\documentclass[
  aps,
  prb,
  reprint,
  superscriptaddress,
  floatfix,
  longbibliography
]{revtex4-2}
\usepackage[T1]{fontenc}
\usepackage[utf8]{inputenc}
\usepackage{amsmath,amssymb}
\usepackage{braket}
\usepackage{graphicx}
\usepackage{hyperref}
\usepackage{xcolor}
\newcommand{\exciting}{\texttt{exciting}}
\newcommand{\ie}{i.e., }
\newcommand{\eg}{e.g., }

\begin{document}


\title{Fredholm approximations to the Bethe-Salpeter equation with
spherical-Bessel kernels for Wannier-Mott excitons}

\author{Christian Tantardini}
\email{christiantantardini@ymail.com}
\affiliation{Center for Integrative Petroleum Research, King Fahd University of Petroleum and Minerals, Dhahran 31261, Saudi Arabia}

\author{Benedikt Maurer}
\affiliation{Department of Physics and Center for the Science of Materials Berlin, Humboldt-Universit\"at zu Berlin, 12489 Berlin, Germany}

\author{Claudia Draxl}
\email{claudia.draxl@physik.hu-berlin.de}
\affiliation{Department of Physics and Center for the Science of Materials Berlin, Humboldt-Universit\"at zu Berlin, 12489 Berlin, Germany}

\date{\today}

\begin{abstract}
We develop a deterministic radial approximation to the static
Bethe-Salpeter equation (BSE), designed for rapid preliminary
screening of Wannier-Mott exciton binding across direct-gap bulk
materials. The approximation replaces the electron-hole transition
problem by small radial eigenvalue problems that require no
stochastic sampling. Starting from the four-point BSE and its
matrix representation in a Bloch electron-hole basis, we retain
one valence and one conduction band in the effective-mass limit
within the Tamm-Dancoff approximation. We introduce a central
projection of the screened direct interaction and omit bare
electron-hole exchange. The resulting model applies to spatially
extended excitons whose envelopes vary slowly on the scale of
the unit cell, provided that the effective-mass description and
central screened interaction adequately represent the exciton
of interest. A partial-wave decomposition separates the
angular-momentum channels and expresses their interaction kernels
as spherical-Bessel transforms of the real-space screened Coulomb
interaction. Nystr\"{o}m discretization produces real symmetric
radial matrices suitable for standard eigensolvers. Their dimension
is set by the radial grid independently of the Brillouin-zone
sampling used to calculate screening. We implement the formulation
in the all-electron, full-potential code \exciting. For bulk GaAs,
the compact radial problem reproduces the corresponding conventional
matrix-BSE binding energy within \(1~\mathrm{meV}\) at matched
sampling. Further refinement achieves finite-mesh stability within
the same tolerance on denser screening meshes without increasing
the radial matrix size.
\end{abstract}

\maketitle

\section{Introduction}

Neutral excitations in solids, in particular bound electron-hole pairs (excitons), give rise to prominent features in the optical spectra of semiconductors, insulators, and low-dimensional materials. Their reliable quantitative description generally requires an excplicit treatment of the electron-hole interaction beyond the independent-particle picture and commonly uses adiabatic approximations to time-dependent density-functional theory (TDDFT). The most widely used approach today is many-body perturbation theory (MBPT) in the $GW$+Bethe-Salpeter equation (BSE) framework, where quasiparticle energies obtained from the $GW$ approach are combined with a BSE for the two-particle Green's function\cite{OnidaRMP2002,StrinatiRNC1988,ReiningBSELecture2016,Blase_2020}. This scheme has become a \textit{de~facto} standard for calculating excitons and optical spectra in bulk and low-dimensional systems\cite{OnidaRMP2002,SottilePRB2007,FuchsPRB2008,QuintelaPSSB2022}, and has also been successfully applied to molecules and complex environments\cite{Blase_2020,ForsterJCTC2022}.

Historically, the BSE was introduced in relativistic quantum field theory to describe two-particle bound states in quantum electrodynamics, with an interaction kernel constructed from propagators and interaction vertices\cite{BetheSalpeter1951,salpeter1951salpeter}.
In condensed-matter and quantum-chemical applications, the same formal structure is typically recast into an equal-time (Salpeter-like) equation adapted to electron-hole excitations in crystalline or finite systems\cite{OnidaRMP2002,Blase_2020,ReiningBSELecture2016,LuchaInstantaneous2000,CungAnnPhys1976}.
In all these settings, the BSE is fundamentally an integral equation for a two-particle amplitude with a nonlocal interaction kernel. In practical solid-state calculations, however, the BSE is almost always implemented in matrix form as an effective electron-hole Hamiltonian acting in the space of Bloch transitions\cite{OnidaRMP2002,StrinatiRNC1988,ReiningBSELecture2016}.
The diagonal part contains quasiparticle transition energies, while the off-diagonal elements arise from the screened direct (attractive) and bare exchange (repulsive) Coulomb interactions\cite{StrinatiRNC1988,AlbrechtPRL1998,RohlfingLouiePRB2000}.
This formulation has been implemented in many \emph{ab initio} codes and applied to bulk semiconductors, molecular crystals, nanostructures, and two-dimensional materials,\cite{OnidaRMP2002,Sangalli_2019,Gonze_2016,Gonze_2020,Romero2020}.
Since its computational cost grows rapidly with the number of bands and $\mathbf{k}$-points, and storing and constructing the full kernel can become a serious bottleneck, which has triggered intensive work on more efficient solvers and kernel representations: Haydock-type time-evolution schemes\cite{Haydock1980,Gruning2009}, reduced-scaling and tensor-based diagonalization \cite{SchmidtPRB2003,FuchsPRB2008,BennerMPC2016}, finite-field  methods \cite{PhysRevLett.122.237402} and stochastic approaches \cite{Bradbury_2022}, as well as low-rank and interpolative kernel decompositions\cite{henneke2020fast,BennerMPC2016}.
Recent developments include real-time BSE propagation \cite{MarekJCTC2025}, GPU-accelerated BSE implementations in planewave codes \cite{WestBSEGPU2024}, data-driven low-rank kernels in time-dependent $GW$-based schemes \cite{HouNPJCM2025}, and subspace or model-reduction strategies tailored to specific excitations\cite{QiuPRB2021}. Nevertheless, these advances do not alter the underlying mathematical character of the BSE as an integral equation for bound excitons. 

From a mathematical perspective, once momentum space is restricted to a bounded domain and the interaction kernel is regularized so that it is square-integrable on the corresponding product domain, the associated integral operator is Hilbert-Schmidt and therefore compact. This compactness provides a basis for isolating bound-state eigenvalues and constructing systematic finite-rank approximations \cite{KressLIE,HackbuschIE,AtkinsonSIAM1967,DelvesMohamedIE}.
Such Fredholm techniques have a long history in relativistic Bethe-Salpeter studies: early work by Graves-Morris, Broido and Taylor, Greenman, Bauhoff, and others explicitly constructed ``Fredholm approximations'' by projecting the kernel onto finite-dimensional subspaces and analyzing convergence of the resulting matrix problems \cite{GravesMorrisPRL1966,BroidoTaylorJMP1969,GreenmanNCB1970,BauhoffZNA1974,cutkosky1954solutions,kaufmann1969numerical}. These ideas, however, have not been systematically brought into the modern solid-state $GW$+BSE framework.
In parallel, full matrix-BSE solvers have been connected to simpler effective descriptions in terms of Wannier-Mott excitons and effective-mass Hamiltonians\cite{FuchsPRB2008,QuintelaPSSB2022,Blase_2020}.
In the Wannier-Mott regime, where the exciton radius spans many unit cells, the exciton behaves as a hydrogenic quasiparticle whose relative motion is governed by a screened Coulomb potential with material-dependent effective masses and dielectric constant \cite{QuintelaPSSB2022}. In the continuum limit, this leads to a Wannier equation for the exciton envelope in real or momentum space, and the effective-mass BSE becomes explicitly a three-dimensional integral equation \cite{GreenmanNCB1970,LuchaInstantaneous2000,CungAnnPhys1976}. In this language, the link to Fredholm theory follows naturally.



In this work, we develop a radial Fredholm approximation to the
solid-state BSE for rapid preliminary screening of direct-gap bulk
materials through their Wannier-Mott exciton binding energies.
Our strategy reduces the physical description to small radial
eigenvalue problems while retaining material-dependent effective
masses and microscopic screening. The radial matrices are solved
deterministically, without the sampling uncertainty associated
with stochastic BSE formulations \cite{Bradbury_2022}.
Their dimension is controlled independently of the Brillouin-zone
sampling used to calculate the screened Coulomb interaction,
allowing denser screening meshes without enlarging the exciton
matrices. We implement the formulation in the all-electron,
full-potential code \exciting\ and assess its accuracy and
finite-mesh stability for bulk GaAs through comparison with
the conventional matrix BSE at identical sampling and systematic
mesh refinement.

The remainder of the paper is organized as follows. Section~\ref{sec:mb-bse-solids} reviews the BSE and its matrix representation in a Bloch electron-hole basis. Section~\ref{sec:bse-fred} derives the momentum-space Wannier equation and its Fredholm formulation. Section~\ref{sec:PartialWaves} introduces the partial-wave decomposition and the spherical-Bessel representation of the central kernel. Section~\ref{sec:Numerics} describes the Nystr\"{o}m discretization and matrix symmetrization. Section~\ref{sec:res} validates the implementation against the conventional matrix BSE and shows its numerical stability for the example of bulk GaAs. Finally, Section~\ref{sec:conc} summarizes the results
and their range of applicability.

\section{Bethe-Salpeter equation for solids}
\label{sec:mb-bse-solids}

\subsection{Two-particle Green's function and four-point BSE}

We consider an interacting many-electron system described by a microscopic Hamiltonian $\hat{H}$, with ground state $\ket{\mathrm{GS}}$ and ground-state energy $E_{\mathrm{GS}}$.
The central object of the Bethe-Salpeter formalism is the time-ordered two-particle (electron-hole) Green's function
\begin{align}
  L(12,1'2')
  &= - i \, \langle \mathrm{GS} | \,
      T \, \hat{\psi}(1)\hat{\psi}^\dagger(2)
           \hat{\psi}(2')\hat{\psi}^\dagger(1')
      \, | \mathrm{GS} \rangle,
  \label{eq:def_L}
\end{align}
where we use compressed notations $1 \equiv (\mathbf{r}_1,t_1,\sigma_1)$, $2 \equiv (\mathbf{r}_2,t_2,\sigma_2)$, etc., with $\sigma$ being the spin index, and $T$ denoting time ordering. The field operators $\hat{\psi}$ and $\hat{\psi}^\dagger$ are taken in the Heisenberg picture. Throughout this work, we employ zero-temperature, real-time, and time-ordered Green's functions. The function $L$ describes the propagation of an electron-hole pair and contains, in particular, the information on neutral excitations and optical spectra \cite{StrinatiRNC1988,OnidaRMP2002}.

The corresponding one-particle Green's function is
\begin{align}
  G(1,1')
  &= - i \, \langle \mathrm{GS} | \,
    T \, \hat{\psi}(1) \hat{\psi}^\dagger(1') \, |\mathrm{GS}\rangle,
\end{align}
which is the fundamental quantity of $GW$ theory \cite{OnidaRMP2002}. Within many-body Green's-function theory, the equations of motion for $G$ and $L$, together with Hedin's equations for the self-energy $\Sigma$, lead to a Dyson-like integral equation for the two-particle Green's function. In time-ordered four-point form one obtains the BSE
\begin{align}
  L(12,1'2')
  &= L_0(12,1'2') \nonumber \\
  &\quad + \int d(3456)\,
     L_0(13,1'4)\,
     \Xi(35,46)\,
     L(52,62'),
  \label{eq:BSE_fourpoint}
\end{align}
where the shorthand $\int d(3456) \equiv \int d3\, d4\, d5\, d6$ denotes integration over the spatial and temporal coordinates of $3,4,5,6$ and summation over spin indices.
The quantity
\begin{align}
  L_0(12,1'2')
  &= - i \, G(1,2') \, G(2,1')
\end{align}
is the independent-quasiparticle (``noninteracting'') two-particle correlation function obtained when the electron-hole interaction is neglected so that the two propagators factorize. The kernel
\begin{align}
  \Xi(35,46)
  &= i \,\frac{\delta \Sigma(3,4)}{\delta G(6,5)}
\end{align}
is the irreducible electron-hole interaction, defined as the functional derivative of the self-energy with respect to $G$ \cite{StrinatiRNC1988,OnidaRMP2002}.
Equation~\eqref{eq:BSE_fourpoint} is the BSE in its most general four-point form and resumes repeated electron-hole scatterings to all orders.

For optical excitations, we focus on a time-translation invariant system and work in the frequency domain. Introducing relative and center-of-mass times and Fourier transforming with respect to the total time, one can write
\begin{align}
  L(12,1'2';\omega)
  &= L_0(12,1'2';\omega) \nonumber \\
  &+ \int d(3456)\,
     L_0(13,1'4;\omega)\, \nonumber \\
     & \phantom{--} \Xi(35,46;\omega)\,
     L(52,62';\omega).
  \label{eq:BSE_freq}
\end{align}
The reducible polarization (density-density response function) is obtained by contracting $L$ over two of its arguments.
A convenient definition at the two-point level is
\begin{align}
  P(12;\omega)
  &= - i \, L(12,1^+ 2;\omega),
  \label{eq:P_from_L}
\end{align}
where $1^+$ denotes a time infinitesimally later than $1$ to enforce proper
time ordering \cite{StrinatiRNC1988,OnidaRMP2002}.
From the $\mathbf{Q}\to 0$ limit of $P$, where $\mathbf{Q}$ denotes the external wave vector, one extracts the macroscopic dielectric function and optical
spectra.

We use the standard \(GW\)+BSE kernel, neglecting the contribution
proportional to \(\delta W/\delta G\). This assumes that creating
the electron--hole excitation changes the background screening
only weakly \cite{OnidaRMP2002}.
We also evaluate the screened interaction at zero frequency,
\(W(\omega=0)\). This neglects dynamical screening and is appropriate
when the relevant electron--hole energy transfers are small compared
with the characteristic screening energies
\cite{RohlfingLouiePRB2000}.

\subsection{Electron-hole basis and matrix BSE for excitons}

We now express Eq.~\eqref{eq:BSE_freq} in a Bloch basis appropriate for crystalline solids. Let $\{\varphi_{\nu\mathbf{k}}(\mathbf{r})\}$ be Bloch eigenstates with band index $\nu$ and crystal momentum $\mathbf{k}$,
\begin{align}
  \hat{h}_{\text{qp}} \, \varphi_{\nu\mathbf{k}}(\mathbf{r})
  &= \varepsilon_{\nu\mathbf{k}} \, \varphi_{\nu\mathbf{k}}(\mathbf{r}),
\end{align}
where $\hat{h}_{\text{qp}}$ is an effective single-particle Hamiltonian (\eg a quasiparticle $GW$ Hamiltonian) and $\varepsilon_{\nu\mathbf{k}}$ the corresponding quasiparticle energies \cite{OnidaRMP2002,RohlfingLouiePRB2000}. The bands are partitioned into valence ($v$) and conduction ($c$) sets. At vanishing total crystal momentum $\mathbf{Q}\approx 0$ (relevant for optical absorption), the natural electron-hole basis functions are
\begin{align}
  \ket{v c \mathbf{k}}
  &\equiv
  \hat{a}^\dagger_{c\mathbf{k}} \hat{a}^{\phantom{\dagger}}_{v\mathbf{k}} \ket{\mathrm{GS}},
\end{align}
where $\hat{a}^\dagger_{c\mathbf{k}}$ creates an electron in a conduction band and $\hat{a}^{\phantom{\dagger}}_{v\mathbf{k}}$ annihilates an electron in a valence band. The set $\{\ket{v c \mathbf{k}}\}$ spans the subspace of single electron-hole excitations at fixed total momentum.

The eigenstates \(\ket{\lambda}\) of the BSE, which encompass both bound excitons and unbound electron-hole excitations, are expanded in the electron-hole transition basis as
\begin{align}
  \ket{\lambda}
  &= \sum_{v c \mathbf{k}} A^{\lambda}_{v c \mathbf{k}}\,
       \hat{a}^\dagger_{c\mathbf{k}} \hat{a}^{\phantom{\dagger}}_{v\mathbf{k}} \ket{\mathrm{GS}},
  \label{eq:exciton_expansion}
\end{align}
where \(\lambda\) labels a BSE solution, \(A^{\lambda}_{vc\mathbf{k}}\) are the
corresponding BSE amplitudes, and \(E_{\lambda}\) is its excitation energy. 
Inserting this expansion into the equations of motion for $L$,
or equivalently into a variational principle for the polarization
operator, gives the matrix representation of the BSE
\cite{StrinatiRNC1988,OnidaRMP2002}.
We assume a static interaction kernel and use the Tamm--Dancoff
approximation (TDA), which neglects the coupling between resonant
and antiresonant (de-excitation) blocks.
The TDA is appropriate when this coupling is weak compared with
the energy separation between the two sectors
\cite{OnidaRMP2002,RohlfingLouiePRB2000}.
Under these approximations, the matrix BSE eigenvalue problem reads
\begin{align}
  \sum_{v' c' \mathbf{k}'}
  H^{\text{BSE}}_{v c \mathbf{k}, v' c' \mathbf{k}'}
  \, A^{\lambda}_{v' c' \mathbf{k}'}
  &= E_{\lambda} \, A^{\lambda}_{v c \mathbf{k}},
  \label{eq:BSE_eigen_matrix}
\end{align}
where
\begin{align}
  H^{\text{BSE}}_{v c \mathbf{k}, v' c' \mathbf{k}'}
  &= \left( \varepsilon_{c\mathbf{k}} - \varepsilon_{v\mathbf{k}} \right)
     \delta_{vv'} \delta_{cc'} \delta_{\mathbf{k}\mathbf{k}'} \nonumber \\
  &\quad + H^{\text{dir}}_{v c \mathbf{k}, v' c' \mathbf{k}'}
   + H^{\text{exc}}_{v c \mathbf{k}, v' c' \mathbf{k}'}
  \label{eq:BSE_Hamiltonian}
\end{align}
is the effective electron--hole Hamiltonian used in first-principles
calculations of optical spectra and exciton binding energies.

The first term in Eq.~\eqref{eq:BSE_Hamiltonian} is the independent-particle transition energy, given by the difference of quasiparticle band energies. The interaction terms read, in real space,
\begin{align}
  H^{\mathrm{dir}}_{vc\mathbf{k},v'c'\mathbf{k}'}
  &=
  -\int d^3r\,d^3r'\,
  \varphi_{c\mathbf{k}}^*(\mathbf{r})
  \varphi_{c'\mathbf{k}'}(\mathbf{r})
  W(\mathbf{r},\mathbf{r}')
  \nonumber\\
  &\qquad\qquad\times
  \varphi_{v\mathbf{k}}(\mathbf{r}')
  \varphi_{v'\mathbf{k}'}^*(\mathbf{r}'),
  \label{eq:K_dir_real}
  \\
  H^{\mathrm{exc}}_{vc\mathbf{k},v'c'\mathbf{k}'}
  &=
  \int d^3r\,d^3r'\,
  \varphi_{c\mathbf{k}}^*(\mathbf{r})
  \varphi_{v\mathbf{k}}(\mathbf{r})
  v(\mathbf{r},\mathbf{r}')
  \nonumber\\
  &\qquad\qquad\times
  \varphi_{c'\mathbf{k}'}(\mathbf{r}')
  \varphi_{v'\mathbf{k}'}^*(\mathbf{r}').
  \label{eq:K_exc_real}
\end{align}
Here \(W(\mathbf{r},\mathbf{r}')\) is the statically screened Coulomb interaction and \(v(\mathbf{r},\mathbf{r}')\) is the long-range part of the bare Coulomb interaction \cite{StrinatiRNC1988,OnidaRMP2002}. The direct term is attractive, whereas the exchange term is repulsive. In the spin convention used for the matrix-BSE calculations, the exchange contribution is retained in the singlet channel and vanishes in the triplet channel. This is why the triplet calculation provides the closest like-for-like reference for the screened-direct Fredholm model. 
Equation~\eqref{eq:BSE_eigen_matrix} thus provides the link between the many-body Green's-function formulation of the BSE and the practical matrix eigenvalue problem solved in most first-principles BSE implementations for solids.

\section{From matrix BSE to a Fredholm integral equation}
\label{sec:bse-fred}

In this section, we provide a derivation of a Fredholm integral formulation for the exciton envelope from the matrix BSE. Starting from a Bloch electron-hole basis, we introduce the one-valence-band/one-conduction-band, effective-mass, and screened-direct approximations that lead to the momentum-space Wannier equation. We then recast this equation in Fredholm form, specify the regularized finite-domain operator used in the numerical treatment, and extend the formulation to anisotropic effective masses. 

\subsection{Effective-mass and Wannier limit}

We now reduce Eq.~\eqref{eq:BSE_eigen_matrix} to the Wannier-Mott regime, where the relevant electron-hole transitions lie close to the band extrema and the amplitudes \(A^{\lambda}_{vc\mathbf{k}}\) vary smoothly with \(\mathbf{k}\) \cite{StrinatiRNC1988,QuintelaPSSB2022}. This motivates replacing the discrete \(\mathbf{k}\)-point sampling by a continuum description in momentum space.
For definiteness, we restrict both the derivation and the numerical validation to direct-gap systems, for which the conduction- and valence-band extrema coincide at a common wave vector $\mathbf{k}_0$. Extending the construction to indirect excitons requires a finite center-of-mass momentum and lies outside the present scope. For the direct-gap case, we retain one valence and one conduction
band near $\mathbf{k}_0$. This selects a single electron--hole
channel and assumes that coupling to other band pairs is weak
\cite{FuchsPRB2008}.
The spatially extended envelope of a Wannier--Mott exciton is
concentrated near the band extrema in momentum space, motivating
parabolic approximations to the band dispersions
\cite{OnidaRMP2002,FuchsPRB2008}.
We first use isotropic effective masses, neglecting directional
variations of the band curvatures, and write
\begin{align}
  \varepsilon_{c\mathbf{k}}
  &\simeq E_c^0 + \frac{|\mathbf{k}-\mathbf{k}_0|^2}{2 m_c},
  &
  \varepsilon_{v\mathbf{k}}
  &\simeq E_v^0 - \frac{|\mathbf{k}-\mathbf{k}_0|^2}{2 m_v},
  \label{eq:EffMassDispersion}
\end{align}
where $E_c^0$ and $E_v^0$ are the band-edge energies, and $m_c$
and $m_v$ are the corresponding effective masses.
The formulation with anisotropic masses is developed in
Sec.~\ref{sec:anisotropic-mass}.
Introducing the shifted momentum
\begin{align}
  \mathbf{p} &= \mathbf{k} - \mathbf{k}_0,
\end{align}
the reduced mass
\begin{align}
  \mu^{-1} &= m_c^{-1} + m_v^{-1},
\end{align}
and the quasiparticle gap $E_g = E_c^0 - E_v^0$, we obtain
\begin{align}
  \varepsilon_{c\mathbf{k}} - \varepsilon_{v\mathbf{k}}
  &\simeq E_g + \frac{p^2}{2\mu},
  \qquad p = |\mathbf{p}|.
  \label{eq:TransitionEnergyEffMass}
\end{align}

The reduced model retains the screened direct attraction responsible for electron--hole binding and omits \(H^{\mathrm{exc}}\).
It therefore describes binding without exchange-induced singlet--triplet splitting \cite{RohlfingLouiePRB2000}.
Here, \(\mathbf Q=0\) refers to the total exciton momentum; the direct kernel can nevertheless depend on the incoming and outgoing relative momenta \(\mathbf p\) and \(\mathbf p'\). We assume that these are its only momentum variables,
\begin{align}
  H^{\text{dir}}_{v c \mathbf{k}, v c \mathbf{k}'}
  &\equiv K^{\text{dir}}(\mathbf{p},\mathbf{p}'),
\end{align}
with $\mathbf{p} = \mathbf{k}-\mathbf{k}_0$ and $\mathbf{p}' = \mathbf{k}'-\mathbf{k}_0$. 
We further neglect the dependence of the screened direct interaction on the absolute position of the electron-hole pair within the crystal. After averaging over lattice-scale variations, the kernel can therefore be expressed in terms of the incoming and outgoing relative momenta,
\begin{align}
  H^{\mathrm{dir}}_{v c \mathbf{k},v c \mathbf{k}'}
  &\equiv
  K^{\mathrm{dir}}(\mathbf{p},\mathbf{p}').
\end{align}
This envelope-level approximation is appropriate in the Wannier-Mott regime, where the exciton envelope varies slowly on the scale of the unit cell. An additional restriction to the central, isotropic component of the interaction will be introduced separately in Section~\ref{sec:PartialWaves} \cite{StrinatiRNC1988,OnidaRMP2002}. Microscopic local-field information is retained only through the screened interaction \(W_{\mathrm{cen}}(r)\), defined in Eq.~\eqref{eq:WcenDefinition} and constructed explicitly in Appendix~\ref{sec:SM_central}.

In this Wannier regime, the exciton amplitudes can be factorized into a periodic part, which is carried by the Bloch functions, and a slowly varying envelope $\Phi_{\lambda}(\mathbf{p})$ according to
\begin{align}
  A^{\lambda}_{v c \mathbf{k}}
  &\equiv \Phi_{\lambda}(\mathbf{p}),
  \label{eq:EnvelopeDefinition}
\end{align}
where $\Phi_{\lambda}$ is a smooth function of the relative momentum~$\mathbf{p}$. Replacing the discrete sum over $\mathbf{k}$ by an integral, $\sum_{\mathbf{k}} \to \int d^3k/(2\pi)^3$, and inserting Eqs.~\eqref{eq:EffMassDispersion}-\eqref{eq:EnvelopeDefinition} into Eq.~\eqref{eq:BSE_eigen_matrix}, we arrive at
\begin{widetext}
\begin{align}
  \left(E_g + \frac{p^2}{2\mu}\right)\Phi_{\lambda}(\mathbf{p})
  + \int \frac{d^3p'}{(2\pi)^3}\,
        K^{\text{dir}}(\mathbf{p},\mathbf{p}')\,
        \Phi_{\lambda}(\mathbf{p}')
  &= E_{\lambda} \, \Phi_{\lambda}(\mathbf{p}),
  \label{eq:WannierMomentumPre}
\end{align}
\end{widetext}
which is the familiar momentum-space Wannier equation for bulk excitons, now obtained directly from the BSE in the effective-mass limit \cite{StrinatiRNC1988,QuintelaPSSB2022}.

\subsection{Momentum-space Wannier equation for the binding energy}

For a bound excitonic state \(\lambda\), the binding energy relative to the quasiparticle continuum edge \(E_g\) is defined as
\begin{align}
  E_B^{\lambda} &\equiv E_g-E_{\lambda}>0.
  \label{eq:binding_energy}
\end{align}

Inserting this into Eq.~\eqref{eq:WannierMomentumPre} and rearranging gives
\begin{align}
  \left( \frac{p^2}{2\mu} + E_B^{\lambda} \right)\Phi_{\lambda}(\mathbf{p})
  &= - \int\frac{d^3p'}{(2\pi)^3}\,
         K^{\text{dir}}(\mathbf{p},\mathbf{p}')\,
         \Phi_{\lambda}(\mathbf{p}'),
  \label{eq:WannierMomentumBinding}
\end{align}
which is a homogeneous integral equation for the exciton envelope $\Phi_{\lambda}(\mathbf{p})$ and its binding energy $E_B^{\lambda}$.
This is the momentum-space analogue of the standard real-space Wannier equation for hydrogenic excitons \cite{StrinatiRNC1988,QuintelaPSSB2022} , with the screened direct kernel $K^{\text{dir}}$ acting as a nonlocal attractive potential.

Equation~\eqref{eq:WannierMomentumBinding} is the momentum-space Wannier equation obtained from the matrix BSE in the effective-mass limit. We now recast it as a Fredholm integral equation.

\subsection{Fredholm integral formulation and operator properties}
\label{subsec:fredholm_domain}

From the viewpoint of integral-equation theory, Eq.~\eqref{eq:WannierMomentumBinding} has the generic structure of a homogeneous Fredholm equation of the second kind,
\begin{align}
  \mathcal{L}_b[\Phi](\mathbf{p})
  &\equiv \left( \frac{p^2}{2\mu} + b \right)\Phi(\mathbf{p}) \nonumber \\
  & + \int\frac{d^3p'}{(2\pi)^3}\,
            K^{\text{dir}}(\mathbf{p},\mathbf{p}')\,\Phi(\mathbf{p}')
   = 0,
  \label{eq:FredholmFormal}
\end{align}
with a positive trial binding energy $b$ as spectral parameter. Equivalently, for any trial binding energy $b>0$, Eq.~\eqref{eq:WannierMomentumBinding} can be divided by $(p^2/2\mu + b)$, thereby recasting it as a fixed-point equation
\begin{align}
  \Phi(\mathbf{p})
  &= \int \frac{d^3p'}{(2\pi)^3}\,\mathcal{K}(\mathbf{p},\mathbf{p}';b)\,
     \Phi(\mathbf{p}'),
  \label{eq:FredholmFixedPoint}
\end{align}
with integral kernel
\begin{align}
  \mathcal{K}(\mathbf{p},\mathbf{p}';b)
  &= -\,\frac{K^{\text{dir}}(\mathbf{p},\mathbf{p}')}
            {\frac{p^2}{2\mu}+b}.
  \label{eq:FredholmKernel3D}
\end{align}
For fixed $b$, Eq.~\eqref{eq:FredholmFixedPoint} is an eigenvalue problem for the linear integral operator $\hat{\mathcal{K}}$ defined by the kernel \eqref{eq:FredholmKernel3D}. The physical binding energies $E_B^{\lambda}$ are those values of $b$ for which $\hat{\mathcal{K}}$ has eigenvalue $1$ and a nontrivial eigenfunction $\Phi_{\lambda}$ \cite{KressLIE,HackbuschIE}.

To define the numerical problem precisely, we restrict momentum space to the bounded domain \(\Omega_\Lambda=\{\mathbf{p}\in\mathbb{R}^3:|\mathbf{p}|<\Lambda\}\), where \(\Lambda\) is chosen such that the exciton envelope is negligible outside the retained interval. We work in \(L^2(\Omega_\Lambda,d^3p)\), with the inner product
\begin{align}
  \langle f,g\rangle
  &=
  \int_{\Omega_\Lambda}d^3p\,
  f^*(\mathbf{p})g(\mathbf{p}).
\end{align}
For a fixed trial energy \(b>0\), suppose that the regularized kernel used in the numerical calculation satisfies
\begin{align}
  \int_{\Omega_\Lambda}d^3p
  \int_{\Omega_\Lambda}d^3p'\,
  \left|
    \mathcal{K}(\mathbf{p},\mathbf{p}';b)
  \right|^2
  &<\infty.
  \label{eq:HilbertSchmidtCondition}
\end{align}
The corresponding finite-section operator is then of Hilbert-Schmidt type and therefore compact \cite{KressLIE,HackbuschIE}.

This statement applies to the regularized kernel that is actually
discretized. Static dielectric screening in a bulk semiconductor generally retains a long-range Coulomb tail; consequently, the unregularized momentum-space Coulomb kernel is weakly singular and is not assumed here to satisfy Eq.~\eqref{eq:HilbertSchmidtCondition}. The central projection and the treatment of the Coulomb head are specified in Appendix~\ref{sec:SM_central}, while the real- and momentum-space cutoffs and finite-section control are given in Appendix~\ref{sec:SM_Lambda}.

Within this regularized finite-section problem, the Nystr\"{o}m discretization generates a finite-rank approximation. Under the regularity conditions of the chosen quadrature, its isolated bound-state eigenvalues and eigenfunctions converge as the quadrature is refined \cite{AtkinsonSIAM1967,DelvesMohamedIE,KressLIE}. In Sections~\ref{sec:PartialWaves} and~\ref{sec:Numerics} we exploit rotational symmetry and spherical Bessel expansions to reduce Eq.~\eqref{eq:FredholmFixedPoint} to a family of one-dimensional radial Fredholm equations, and discuss concrete strategies for constructing and solving such finite-rank approximations.

\subsection{Anisotropic effective masses and rescaled momentum coordinates}
\label{sec:anisotropic-mass}

So far we have assumed an isotropic effective mass, \ie a scalar $\mu$ such that the relative kinetic energy is $p^2/(2\mu)$ with $p = |\mathbf{p}|$. Real materials are often more anisotropic, with different curvatures along different crystallographic directions. In this case, the effective-mass description is naturally written in terms of a mass tensor. A simple way to generalize the analysis is to replace the scalar masses $m_c$ and $m_v$ by symmetric positive-definite tensors $\mathbf{m}_c$ and $\mathbf{m}_v$, and work with the corresponding reduced-mass tensor
\begin{align}
  \boldsymbol{\mu}^{-1}
  &= \mathbf{m}_c^{-1} + \mathbf{m}_v^{-1},
\end{align}
which is again a $3\times 3$ symmetric positive-definite matrix. In the vicinity of the band extrema the quasiparticle dispersion is then approximated as
\begin{align}
  \varepsilon_{c\mathbf{k}} - \varepsilon_{v\mathbf{k}}
  &\simeq E_g + \frac{1}{2}\,
      (\mathbf{k}-\mathbf{k}_0)^\mathsf{T}\,
      \boldsymbol{\mu}^{-1}\,
      (\mathbf{k}-\mathbf{k}_0),
\end{align}
and, in terms of the shifted momentum $\mathbf{p} = \mathbf{k}-\mathbf{k}_0$, the Wannier equation \eqref{eq:WannierMomentumBinding} becomes
\begin{widetext}
\begin{align}
  \left( \frac{1}{2}\,\mathbf{p}^\mathsf{T}\boldsymbol{\mu}^{-1}\mathbf{p}
         + E_B^{\lambda} \right)\Phi_{\lambda}(\mathbf{p})
  &= - \int\frac{d^3p'}{(2\pi)^3}\,
         K^{\text{dir}}(\mathbf{p},\mathbf{p}')\,
         \Phi_{\lambda}(\mathbf{p}').
  \label{eq:WannierMomentumAniso}
\end{align}
\end{widetext}
In the original $\mathbf{p}$ coordinates the kinetic term is no longer rotationally invariant, so the equation does not have the simple ``central-potential'' form used in the isotropic case.

A convenient way to recover an isotropic structure is to perform a linear change of variables that absorbs the mass tensor into the definition of the momentum. Since $\boldsymbol{\mu}^{-1}$ is symmetric positive-definite, it admits a spectral decomposition
\begin{align}
  \boldsymbol{\mu}^{-1}
  &= \mathbf{R}^\mathsf{T}
     \begin{pmatrix}
       \alpha_x & 0 & 0 \\
       0 & \alpha_y & 0 \\
       0 & 0 & \alpha_z
     \end{pmatrix}
     \mathbf{R},
\end{align}
with $\mathbf{R}$ being orthogonal and $\alpha_i>0$.
We introduce a symmetric square root $\mathbf{A}$ such that $\boldsymbol{\mu}^{-1} = \mathbf{A}^\mathsf{T}\mathbf{A}$, for example
\begin{align}
  \mathbf{A}
  &= \mathbf{R}^\mathsf{T}
     \begin{pmatrix}
       \sqrt{\alpha_x} & 0 & 0 \\
       0 & \sqrt{\alpha_y} & 0 \\
       0 & 0 & \sqrt{\alpha_z}
     \end{pmatrix}
     \mathbf{R}.
\end{align}
We then define a rescaled momentum coordinate
\begin{align}
  \mathbf{u} &= \sqrt{\mu_{\text{iso}}}\,\mathbf{A}\,\mathbf{p},
  \label{eq:q-def}
\end{align}
where $\mu_{\text{iso}}$ is an arbitrary reference mass (for instance, an average of the principal components of $\boldsymbol{\mu}$). By construction,
\begin{align}
  \frac{1}{2}\,\mathbf{p}^\mathsf{T}\boldsymbol{\mu}^{-1}\mathbf{p}
  &= \frac{1}{2\mu_{\text{iso}}}\,u^2,
  \qquad u = |\mathbf{u}|.
\end{align}
In the $\mathbf{u}$ variables, Eq.~\eqref{eq:WannierMomentumAniso} takes the form
\begin{align}
  \left( \frac{u^2}{2\mu_{\text{iso}}} + E_B^{\lambda} \right)
    \overline{\Phi}_{\lambda}(\mathbf{u})
  &= - \int \frac{d^3u'}{(2\pi)^3}\,
        \overline{K}^{\text{dir}}(\mathbf{u},\mathbf{u}')\,
        \overline{\Phi}_{\lambda}(\mathbf{u}'),
  \label{eq:WannierMomentumRescaled}
\end{align}
where
\begin{align}
  \overline{\Phi}_{\lambda}(\mathbf{u})
  &= \Phi_{\lambda}(\mathbf{p})\big|_{\mathbf{p}=\mathbf{A}^{-1}\mathbf{u}/\sqrt{\mu_{\text{iso}}}},
  \\
  \overline{K}^{\text{dir}}(\mathbf{u},\mathbf{u}')
  &= \big|\det(\mathbf{S}^{-1})\big|\,
     K^{\text{dir}}(\mathbf{p},\mathbf{p}')
     \Big|_{\mathbf{p}=\mathbf{S}^{-1}\mathbf{u},
            \,\mathbf{p}'=\mathbf{S}^{-1}\mathbf{u}'},
\end{align}
with $\mathbf{S} = \sqrt{\mu_{\text{iso}}}\,\mathbf{A}$ and $d^3p' = \big|\det(\mathbf{S}^{-1})\big|\,d^3u'$. The kinetic term in Eq.~\eqref{eq:WannierMomentumRescaled} is now fully isotropic in $\mathbf{u}$, and the anisotropy is shifted into the transformed kernel $\overline{K}^{\text{dir}}$.

If the screening is approximately isotropic in this rescaled coordinate system (or if one chooses to model $W$ as central in the $\mathbf{u}$-space metric), then $\overline{K}^{\text{dir}}(\mathbf{u},\mathbf{u}')$ depends only on $u$, $u'$ and the angle between them, and Eq.~\eqref{eq:WannierMomentumRescaled} has exactly the same structure as the isotropic equation \eqref{eq:WannierMomentumBinding}, with $p$ replaced by $u$ and $\mu$ by $\mu_{\text{iso}}$. In this representation, all of the partial-wave machinery of Section~\ref{sec:PartialWaves} carries over: we can expand $\overline{\Phi}_{\lambda}(\mathbf{u})$ in spherical harmonics $Y_{\ell m}(\hat{\mathbf{u}})$, decompose the kernel into angular-momentum channels, and arrive at radial Fredholm equations with kernels expressed as spherical-Bessel transforms of an effective central interaction.

The mass rescaling preserves the integral-equation structure, but it does not in general produce decoupled one-dimensional radial equations. Such a radial reduction is available only when the transformed kernel is central; otherwise, different angular-momentum channels remain coupled.

\section{Partial-wave decomposition and spherical Bessel kernels}
\label{sec:PartialWaves}

This section uses rotational symmetry and a central projection of the screened direct interaction to reduce the three-dimensional momentum-space Fredholm equation for the exciton envelope to independent one-dimensional radial equations for each angular-momentum channel. The corresponding radial kernels are then expressed as spherical Bessel transforms of the central screened interaction in real space.

\subsection{Rotational symmetry and angular-momentum channels}

In Eq.~\eqref{eq:WannierMomentumBinding}, the kernel $K^{\text{dir}}(\mathbf{p},\mathbf{p}')$ depends on the full three-dimensional momenta $\mathbf{p}$ and $\mathbf{p}'$. For an interaction that is central in the momentum coordinates
used for the radial reduction,, the kernel depends only on the moduli $p = |\mathbf{p}|$, $p' = |\mathbf{p}'|$ and on the angle between them,
\begin{align}
  K^{\text{dir}}(\mathbf{p},\mathbf{p}')
  &= K^{\text{dir}}\!\big(p,p',\cos\theta\big),
  \quad
  \cos\theta = \hat{\mathbf{p}}\!\cdot\hat{\mathbf{p}}',
\end{align}
with $\hat{\mathbf{p}} = \mathbf{p}/p$ and $\hat{\mathbf{p}}' = \mathbf{p}'/p'$.

Rotational invariance allows the angular and radial degrees of freedom to be
separated. We expand the exciton envelope as
\begin{align}
  \Phi_{\lambda}(\mathbf{p})
  &=
  \sum_{\ell=0}^{\infty}
  \sum_{m=-\ell}^{\ell}
  \Phi^{\lambda}_{\ell m}(p)\,
  Y_{\ell m}(\hat{\mathbf{p}}),
  \label{eq:PhiSphericalExpansion}
\end{align}
and the central direct kernel as
\begin{align}
  K^{\mathrm{dir}}(p,p',\cos\theta)
  &=
  \sum_{\ell=0}^{\infty}
  (2\ell+1)\,
  K_\ell(p,p')P_\ell(\cos\theta),
  \label{eq:KLegendreExpansion}
\end{align}
where \(P_\ell(x)\) denotes the Legendre polynomial of degree \(\ell\). Using the orthogonality of the Legendre polynomials, the corresponding partial-wave coefficient is
\begin{align}
  K_\ell(p,p')
  &=
  \frac{1}{2}
  \int_{-1}^{1}d(\cos\theta)\,
  P_\ell(\cos\theta)
  K^{\mathrm{dir}}(p,p',\cos\theta).
  \label{eq:K_l_projection}
\end{align}
Using the addition theorem,
\begin{align}
  P_\ell(\hat{\mathbf{p}}\!\cdot\hat{\mathbf{p}}')
  &=
  \frac{4\pi}{2\ell+1}
  \sum_{m=-\ell}^{\ell}
  Y_{\ell m}(\hat{\mathbf{p}})
  Y_{\ell m}^*(\hat{\mathbf{p}}'),
\end{align}
and the orthonormality of the spherical harmonics, the different \((\ell,m)\) channels decouple. Equation~\eqref{eq:WannierMomentumBinding} then becomes
\begin{align}
  \left(
    \frac{p^2}{2\mu}+E_B^{\lambda}
  \right)
  \Phi^{\lambda}_{\ell m}(p)
  &=
  -\frac{4\pi}{(2\pi)^3}
  \int_0^\infty dp'\,{p'}^2
  K_\ell(p,p')
  \Phi^{\lambda}_{\ell m}(p').
  \label{eq:RadialFredholm}
\end{align}
The factor \(4\pi\) follows explicitly from the addition theorem and must not be absorbed into \(K_\ell\) when the projection in Eq.~\eqref{eq:K_l_projection} is used. For a central interaction, the radial equation is independent of \(m\), and the corresponding eigenvalues are \((2\ell+1)\)-fold degenerate. This reduction is standard in partial-wave treatments of two-body bound states and BSEs \cite{StrinatiRNC1988,
NieuwenhuisTjonFBS1996}.

\subsection{Real-space interaction and spherical Bessel representation}

We now express \(K_\ell(p,p')\) in terms of the central projection of the microscopic screened interaction. The full crystalline interaction \(W(\mathbf{r},\mathbf{r}')\) is generally neither homogeneous nor central. The Fredholm reduction therefore uses the translationally invariant radial component
\begin{align}
  W_{\mathrm{cen}}(r)
  &\equiv
  \mathcal{P}_{\mathrm{cen}}
  \bigl[W(\mathbf{r},\mathbf{r}')\bigr](r),
  \label{eq:WcenDefinition}
\end{align}
where \(\mathcal{P}_{\mathrm{cen}}\) denotes a chosen approximation that
maps the microscopic screened interaction onto a translationally
invariant, central interaction depending only on the electron--hole
separation \(r\). The specific choice used in this work is described
in Appendix~\ref{sec:SM_central}.

The large exciton radius compared with the lattice spacing motivates
averaging over microscopic variations in screening within the unit cell
\cite{OnidaRMP2002,FuchsPRB2008}.
The additional spherical projection assumes that directional variations
of the screened interaction have a small effect on the targeted exciton.
The comparison with the matrix BSE assesses the combined effect of this
projection and the other physical approximations.

The attractive direct kernel is then
\begin{align}
  K^{\mathrm{dir}}(\mathbf{p},\mathbf{p}')
  &=
  -\int d^3r\,
  e^{-i(\mathbf{p}-\mathbf{p}')\cdot\mathbf{r}}\,
  W_{\mathrm{cen}}(r).
  \label{eq:KdirRealSpace}
\end{align}
To relate Eq.~\eqref{eq:KdirRealSpace} to the partial-wave components $K_\ell(p,p')$, we use the Rayleigh expansion of a planewave in spherical harmonics and spherical Bessel functions \cite{MehremAMC2011,DLMF10},
\begin{align}
  e^{i\mathbf{p}\cdot\mathbf{r}}
  &= 4\pi \sum_{\ell=0}^{\infty}\sum_{m=-\ell}^{\ell}
       i^\ell\,j_\ell(pr)\,
       Y_{\ell m}(\hat{\mathbf{p}})\,Y_{\ell m}^*(\hat{\mathbf{r}}),
  \label{eq:PlaneWaveExpansion}
\end{align}
%
%
%
Here $j_\ell(z)$ is the spherical Bessel function of the first kind \cite{DLMF10}.

Applying these expansions to both factors in Eq.~\eqref{eq:KdirRealSpace} and integrating over the angular part of $\mathbf{r}$, we obtain
\begin{align}
  K^{\mathrm{dir}}(\mathbf{p},\mathbf{p}')
  &=
  -4\pi
  \sum_{\ell=0}^{\infty}
  \sum_{m=-\ell}^{\ell}
  Y_{\ell m}(\hat{\mathbf{p}})
  Y_{\ell m}^*(\hat{\mathbf{p}}')
  V_\ell(p,p').
  \label{eq:KdirPartialYlm}
\end{align}
with the \emph{radial} kernel
\begin{align}
  V_\ell(p,p')
  &=
  4\pi
  \int_0^\infty dr\,r^2\,
  j_\ell(pr)j_\ell(p'r)
  W_{\mathrm{cen}}(r).
  \label{eq:VlBessel}
\end{align}
Comparing Eq.~\eqref{eq:KdirPartialYlm} with the Legendre expansion \eqref{eq:KLegendreExpansion} and using the addition theorem, we identify
\begin{align}
  K_\ell(p,p') &= -V_\ell(p,p').
\end{align}
Since the radial equation is independent of \(m\), we henceforth suppress that index and write \(\Phi_{\lambda\ell}(p)\). Inserting \(K_\ell=-V_\ell\) into Eq.~\eqref{eq:RadialFredholm} gives
\begin{align}
  \left(
    \frac{p^2}{2\mu}+E_B^{\lambda}
  \right)
  \Phi_{\lambda\ell}(p)
  &=
  \frac{4\pi}{(2\pi)^3}
  \int_0^\infty dp'\,{p'}^2
  V_\ell(p,p')
  \Phi_{\lambda\ell}(p').
  \label{eq:RadialFredholmVl}
\end{align}
The transform \(V_\ell\) is defined without an attractive minus sign; the attractive sign appears explicitly in the Hamiltonian below.

Equation~\eqref{eq:RadialFredholmVl} is the key result of this section: once rotational symmetry is used, the original three-dimensional Fredholm problem \eqref{eq:WannierMomentumBinding} reduces to a family of one-dimensional radial integral equations whose kernels are determined by spherical-Bessel transforms of the central screened interaction $W_{\mathrm{cen}}(r)$. In the next section, we use this structure as the starting point for concrete numerical Fredholm approximations and their implementation.

\section{Fredholm approximation and numerical implementation}
\label{sec:Numerics}

In this section, the one-dimensional radial Fredholm equations are converted into finite-dimensional eigenvalue problems suitable for numerical solution and implemented in \exciting. We first define the regularized radial operator and establish its symmetry properties, and then apply Nystr\"{o}m discretization and a weight rescaling to obtain real symmetric matrices compatible with standard eigensolvers.

\subsection{Radial Fredholm operator and symmetrization}

The partial-wave decomposition of Section~\ref{sec:PartialWaves} turns the three-dimensional Fredholm equation \eqref{eq:WannierMomentumBinding} into a family of one-dimensional radial equations of the form \eqref{eq:RadialFredholmVl}, where the kernel $V_\ell$ is given by the spherical Bessel transform of the real-space interaction, Eq.~\eqref{eq:VlBessel}.

For each angular momentum $\ell$ we define the radial operator
\begin{align}
  \bigl(\hat{\mathcal{H}}_\ell f\bigr)(p)
  &=
  \frac{p^2}{2\mu}\,f(p)
  -
  \frac{4\pi}{(2\pi)^3}
  \int_0^\infty dp'\,{p'}^2
  V_\ell(p,p')f(p').
  \label{eq:RadialOperatorDef}
\end{align}
We denote the radial eigenvalue relative to the quasiparticle
continuum edge by \(\mathcal{E}_{\lambda}\).
For a bound state, \(\mathcal{E}_{\lambda}=-E_B^{\lambda}<0\).
Bound excitons in channel \(\ell\) therefore satisfy
\begin{align}
  \hat{\mathcal{H}}_\ell\Phi_{\lambda\ell}
  &=
  -E_B^{\lambda}\Phi_{\lambda\ell}.
\end{align}
On the finite interval \(0\le p\le\Lambda\), we use the same regularized finite-section kernel introduced in Section~\ref{sec:bse-fred}. Requiring
\begin{align}
  \int_0^\Lambda dp\,p^2
  \int_0^\Lambda dp'\,{p'}^2
  \left|V_\ell(p,p')\right|^2
  <\infty ,
\end{align}
the interaction part of \(\hat{\mathcal H}_\ell\) is then of Hilbert-Schmidt type and therefore compact. This statement concerns the regularized finite-section interaction, not the unregularized three-dimensional Coulomb kernel. The bound states correspond to isolated eigenvalues of the full radial Hamiltonian and can be approximated by finite-rank Nystr\"{o}m sequences.

The finite-section rationale, boundary-tail diagnostic, and fixed radial
parameters used in the GaAs calculations are given in Appendix~\ref{sec:SM_Lambda}.

We work with the radial scalar product
\begin{align}
  \langle f,g\rangle_\ell
  &= \int_0^\Lambda dp\,p^2\,f^*(p)\,g(p),
\end{align}
for which $\hat{\mathcal{H}}_\ell$ is self-adjoint when $V_\ell(p,p')$ is symmetric.
At the discrete level, we enforce this symmetry by a simple rescaling of grid values and quadrature weights. 
The radial kernels $V_\ell(p_i,p_j)$ are evaluated by the tabulated spherical-Bessel quadrature and symmetry-preserving kernel assembly described in Appendix~\ref{sec:FastKernels}.

\subsection{Nystr{\"o}m discretization of the radial equation}

To obtain a finite matrix problem, we use a Nystr{\"o}m discretization of Eq.~\eqref{eq:RadialFredholmVl} based on a quadrature rule on $[0,\Lambda]$. We choose $\{p_i\}_{i=1}^N$ as the radial grid with weights $\{w_i\}_{i=1}^N$,
\begin{align}
  \int_0^\Lambda f(p)\,dp \;\approx\; \sum_{i=1}^N w_i f(p_i).
\end{align}
The present implementation uses a uniform radial momentum grid. Further numerical details, including the grid size and momentum cutoff, are documented in Appendices~\ref{sec:SM_Lambda}
and~\ref{sec:FastKernels}.

Evaluating Eq.~\eqref{eq:RadialFredholmVl} at \(p=p_i\) and applying the
quadrature gives
\begin{align}
  \left(
    \frac{p_i^2}{2\mu}+E_B^{\lambda}
  \right)
  \Phi_{\lambda\ell}(p_i)
  &\approx
  \frac{4\pi}{(2\pi)^3}
  \sum_{j=1}^{N}
  p_j^2w_j
  V_\ell(p_i,p_j)
  \Phi_{\lambda\ell}(p_j).
\end{align}
Introducing
\(\Phi_{\ell,i}\equiv\Phi_{\lambda\ell}(p_i)\) and
\begin{align}
  \bigl[\mathbf{D}^{(\ell)}\bigr]_{ij}
  &=
  \delta_{ij}\frac{p_i^2}{2\mu},
  \\
  \bigl[\mathbf{U}^{(\ell)}\bigr]_{ij}
  &=
  \frac{4\pi}{(2\pi)^3}
  p_j^2w_jV_\ell(p_i,p_j),
\end{align}
we obtain
\begin{align}
  \left(
    \mathbf{D}^{(\ell)}-\mathbf{U}^{(\ell)}
  \right)
  \boldsymbol{\Phi}_\ell
  &=
  -E_B^{\lambda}\boldsymbol{\Phi}_\ell.
  \label{eq:MatrixFredholm}
\end{align}
Thus, if
\(\mathbf{H}^{(\ell)}\boldsymbol{\Phi}_\ell
=\mathcal{E}_{\lambda}\boldsymbol{\Phi}_\ell\),
with
\(\mathbf{H}^{(\ell)}
=\mathbf{D}^{(\ell)}-\mathbf{U}^{(\ell)}\)
and \(\mathcal{E}_{\lambda}<0\), then \(E_B^{\lambda}=-\mathcal{E}_{\lambda}\).

To use standard solvers for Hermitian matrices, we symmetrize the matrix by absorbing the measure $p^2dp$ and the quadrature weights into the unknowns. Defining
\begin{align}
  \widetilde{\Phi}_{\ell,i}
  &=
  \sqrt{w_i}\,p_i\,\Phi_{\lambda\ell}(p_i),
  \\
  \widetilde{U}^{(\ell)}_{ij}
  &=
  \frac{4\pi}{(2\pi)^3}
  \sqrt{w_iw_j}\,p_ip_j
  V_\ell(p_i,p_j),
\end{align}
the symmetrized Hamiltonian is
\begin{align}
  \widetilde{H}^{(\ell)}_{ij}
  &=
  \delta_{ij}\frac{p_i^2}{2\mu}
  -
  \widetilde{U}^{(\ell)}_{ij}.
\end{align}
It is real symmetric when \(V_\ell(p_i,p_j)=V_\ell(p_j,p_i)\), and
\begin{align}
  \widetilde{\mathbf H}^{(\ell)}
  \widetilde{\boldsymbol{\Phi}}_\ell
  &=
  \mathcal{E}_{\lambda}\widetilde{\boldsymbol{\Phi}}_\ell
\end{align}
has the same eigenvalues as Eq.~\eqref{eq:MatrixFredholm}. This formulation allows for using the LAPACK backend in \exciting\ to obtain the solution \cite{laug}.

\section{Numerical validation and application to bulk GaAs}
\label{sec:res}

We use bulk GaAs to assess the radial Fredholm formulation in the direct-gap Wannier-Mott regime. 
Both the conventional matrix BSE and the reduced calculation use a static interaction kernel, the Tamm-Dancoff approximation, and one valence and one conduction band.
The reduced formulation additionally approximates the band dispersions near the gap by effective-mass expressions, neglects bare electron-hole exchange, and replaces the microscopic screened interaction by a translationally invariant central interaction. 
The formulation is applicable to Wannier-Mott excitons in direct-gap systems within the stated approximations and is not specific to GaAs.
The mass rescaling and partial-wave decomposition are transformations within the chosen model, whereas the finite-domain restriction and Nystr\"{o}m discretization are numerical approximations.

The assessment addresses three questions. First, we compare the lowest Fredholm and conventional matrix-BSE eigenvalues using the same first-principles input and identical Brillouin-zone meshes.
This comparison assesses the combined effect of the additional physical approximations and the radial implementation. 
Second, we examine whether the screened interaction and radial eigenfunctions stabilize as the Brillouin-zone sampling is refined. 
Third, we quantify the remaining sampling dependence through the change in the Fredholm binding energy between successive meshes. 
The latter two tests assess numerical stability within the reduced model; they do not establish the accuracy of its physical approximations.

\begin{figure}[!t]
  \centering
  \includegraphics[width=\columnwidth]
  {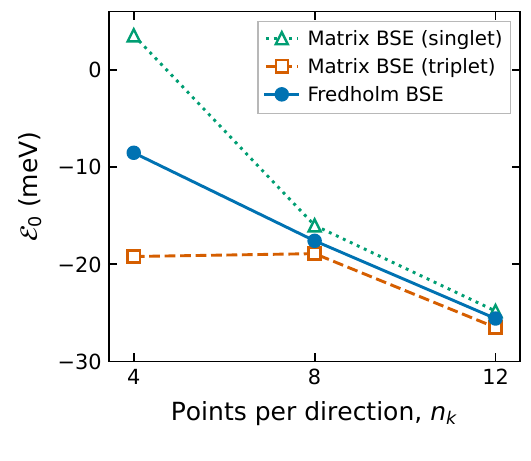}
 \caption{Validation of the radial Fredholm BSE against the conventional singlet and triplet matrix BSE for bulk GaAs. The lowest energies \(\mathcal{E}_0\) relative to the quasiparticle continuum edge are shown for \(n_k\times n_k\times n_k\) Brillouin-zone meshes with \(n_k=4,8,12\). For bound states, \(\mathcal{E}_0=-E_B\). Lines are guides to the eye.}
  \label{fig:gaas_bse_comparison}
\end{figure}

Throughout this section, \(\mathcal{E}_0<0\) denotes the lowest eigenvalue relative to the quasiparticle continuum edge, and \(E_B=-\mathcal{E}_0>0\) is the corresponding binding energy. We use \(n_k^3\) as shorthand for an
\(n_k\times n_k\times n_k\) Brillouin-zone mesh, where \(n_k\)
is the number of sampling points along each reciprocal-lattice direction. 
For the conventional matrix BSE, \(\mathcal{E}_0^{\mathrm{mat}}\) is obtained
by subtracting the quasiparticle gap from the lowest excitation energy:
\begin{equation}
  \mathcal{E}_0^{\mathrm{mat}}(n_k^3)
  =
  E_0^{\mathrm{mat}}(n_k^3)-E_g(n_k^3),
\end{equation}
where \(E_0^{\mathrm{mat}}(n_k^3)\) is the lowest matrix-BSE excitation
energy on the \(n_k^3\) mesh and \(E_g(n_k^3)\) is the quasiparticle gap used
to construct that BSE Hamiltonian. For the Fredholm calculation, \(\mathcal{E}_0^{\mathrm{F}}\) is the lowest eigenvalue of the radial Hamiltonian in Eq.~\eqref{eq:MatrixFredholm}. Accordingly, Fig.~\ref{fig:gaas_bse_comparison} reports \(\mathcal{E}_0=-E_B\), whereas Fig.~\ref{fig:gaas_delta_convergence} reports the positive absolute change \(\Delta E_B\).

\subsection{Computational setup}
\label{sec:gaas_setup}

Bulk GaAs crystallizes in the zinc-blende structure with the primitive face-centered-cubic lattice vectors
\begin{align}
  \mathbf{a}_1 &= \frac{a}{2}(0,1,1), &
  \mathbf{a}_2 &= \frac{a}{2}(1,0,1), &
  \mathbf{a}_3 &= \frac{a}{2}(1,1,0),
\end{align}
where Ga and As are located at the fractional coordinates \((0,0,0)\) and \((1/4,1/4,1/4)\), respectively. The lattice parameter is $10.673173~\mathrm{bohr}
  \simeq 5.648~\text{\AA}$.

All calculations were performed with the all-electron full-potential code \exciting. The ground-state electronic structure was obtained with the exchange-correlation functional PBEsol \cite{PerdewPBEsol2008}. The basis-set cutoff was set to \(R_{\mathrm{MT}}G_{\max}=7.0\), using  muffin-tin radii of\(R_{\mathrm{MT}}=2.3~\mathrm{bohr}\) for both Ga and As. The angular-momentum cutoffs were \(l_{\max}^{\mathrm{APW}}=8\), \(l_{\max}^{\mathrm{vr}}=8\), and \(l_{\max}^{\mathrm{mat}}=8\), and the interstitial potential and density cutoff was \(G_{\max}^{\mathrm{vr}}=18.0\). A \(12\times12\times12\) \(\mathbf{k}\)-point mesh and 250 empty states were used for the ground-state calculation.

The response calculations were performed with the BSE, using the quasiparticle energies obtained from the preceding \(G_0W_0\) calculation. The dielectric screening was evaluated from the full microscopic dielectric matrix on a mesh of response wave vectors \(\mathbf q\), using 400 empty states and a reciprocal-space cutoff \(G_{\mathbf q,\max}=8.0\).
For every response calculation, the \(\mathbf{k}\)- and
\(\mathbf{q}\)-point meshes had the same density,
\(N_k=N_q=n_k^3\).
The \(\mathbf{k}\)-mesh offset was specified by
\texttt{vkloff="0.013 0.029 0.031"}, with each component
expressed as a fraction of the mesh spacing along the corresponding
reciprocal-lattice direction.
The conventional matrix BSE was evaluated on \(n_k\times n_k\times n_k\) Brillouin-zone meshes with \(n_k\in\{4,8,12\}\). 
For the convergence analysis, the screening and radial Fredholm calculations were additionally performed for \(n_k=16\) and \(20\).
All other numerical parameters were kept fixed.
The benchmark focuses on the selected band-edge exciton channel within a two-band Wannier-Mott description \cite{FuchsPRB2008}.
Both the conventional matrix BSE and the Fredholm model retain the highest valence band and the lowest conduction band, corresponding to \(n_v=22\) and \(n_c=1\) in the \exciting\ input convention.
The comparison therefore assesses the additional effective-mass and central-interaction approximations together with the radial implementation within this common band space.
Both the singlet and triplet matrix-BSE Hamiltonians were solved by direct Hermitian diagonalization.

\begin{figure}[!t]
  \centering
  \includegraphics[width=\columnwidth]{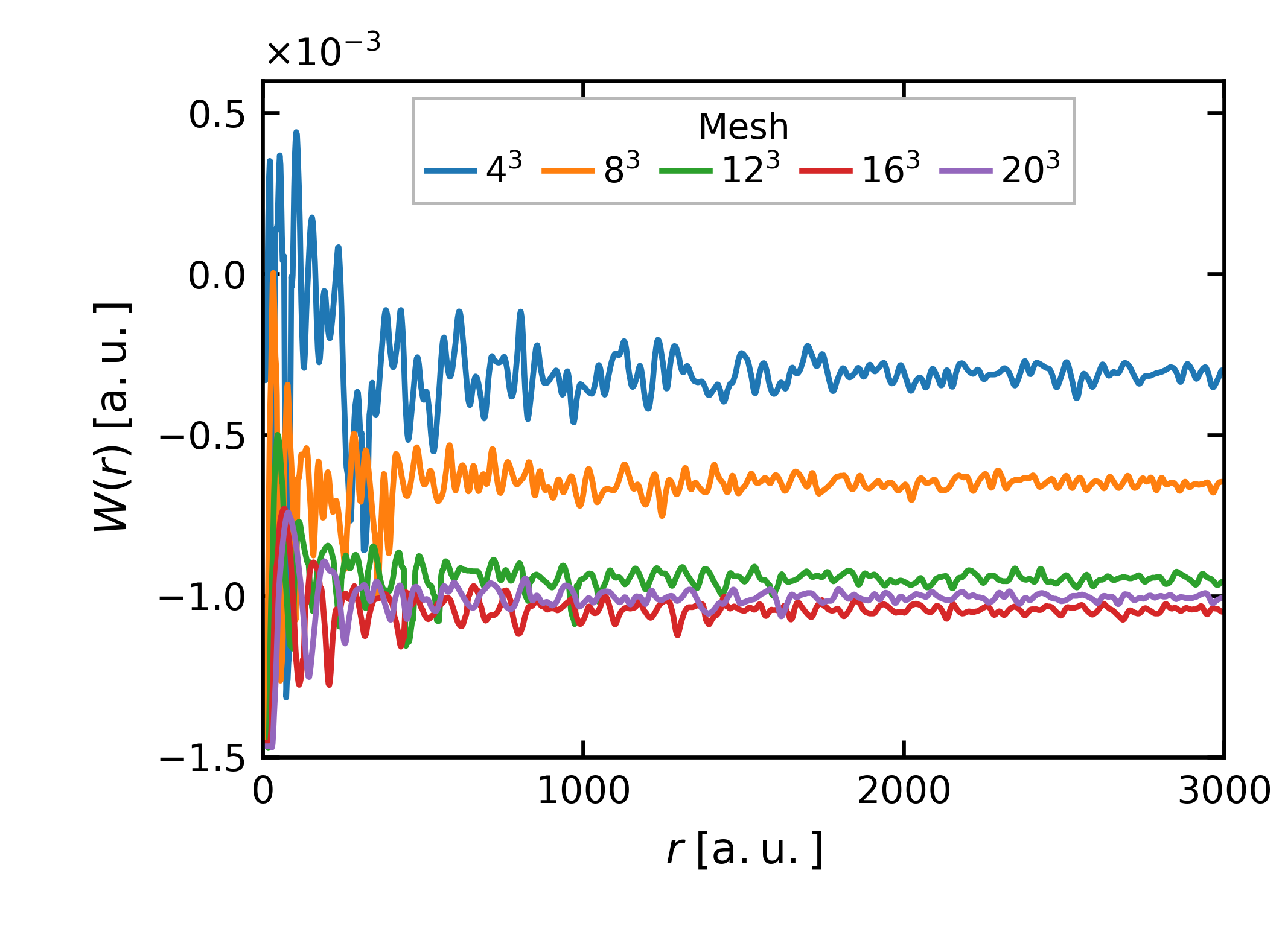}
  \caption{
  Real-space screened Coulomb interaction \(W_{\mathrm{cen}}(r)\), entering the Fredholm BSE (Eq.~\eqref{eq:VlBessel}) shown for bulk GaAs at different Brillouin-zone samplings.  }
  \label{fig:gaas_screened_interaction}
\end{figure}

For the Fredholm calculation, we solve the \(\ell=0\) and \(\ell=1\)
radial equations, corresponding to \(s\)- and \(p\)-like states.
Because the channels decouple for the central interaction
[Eq.~\eqref{eq:RadialFredholm}], omitting higher angular-momentum
channels does not affect the eigenvalues within the retained channels.
The real-space radial grid contained 3000 points and had a maximum radius of \(4000~\mathrm{bohr}\). The radial momentum grid contained 400 points and had a maximum momentum of \(0.02~\mathrm{bohr}^{-1}\).
The additional momentum-cutoff factor was \(p_{\max}^{\mathrm{factor}}=1.05\), and the real-space tail-truncation parameter was 0.7.
The reduced electron-hole mass tensor used in the momentum rescaling described in Section~\ref{sec:anisotropic-mass} was extracted from the quasiparticle dispersions around \(\Gamma\), using three sampling points along each reciprocal-space direction with a spacing \(\Delta k=0.055\) in the reciprocal-coordinate convention of the \exciting\ calculation.
The steps used in the Fredholm implementation are summarized in
Appendix~\ref{sec:SM_implementation}.

\subsection{Validation against the conventional matrix BSE}
\label{sec:gaas_bse_comparison}

We first compare the two formulations on identical Brillouin-zone meshes.
Figure~\ref{fig:gaas_bse_comparison} shows the lowest energies \(\mathcal{E}_0\) relative to the quasiparticle continuum edge for the conventional singlet and triplet matrix BSE and the radial Fredholm formulation on the \(4^3\), \(8^3\), and \(12^3\) meshes.
For a bound state, this energy is \(\mathcal{E}_0=-E_B\). 
The differences between the three results decrease as the sampling is refined.
On the \(12^3\) mesh, the Fredholm value lies between the singlet and triplet results and differs from each by less than \(1~\mathrm{meV}\).

The triplet matrix BSE provides the closest reference because,
like the Fredholm model, it retains the screened direct interaction and omits bare electron-hole exchange. This comparison assesses the effective-mass approximation introduced in Sec.~\ref{sec:bse-fred} and the central projection defined in Eq.~\eqref{eq:WcenDefinition}, together with their numerical
implementation.
The matrix and Fredholm calculations use the same first-principles input and identical Brillouin-zone meshes.
Agreement on the \(12^3\) mesh does not establish convergence with respect to Brillouin-zone sampling; this is examined separately using the denser Fredholm meshes.

\begin{figure*}[!t]
  \centering
  \includegraphics[
    width=\textwidth,
    height=0.9\textheight,
    keepaspectratio
  ]{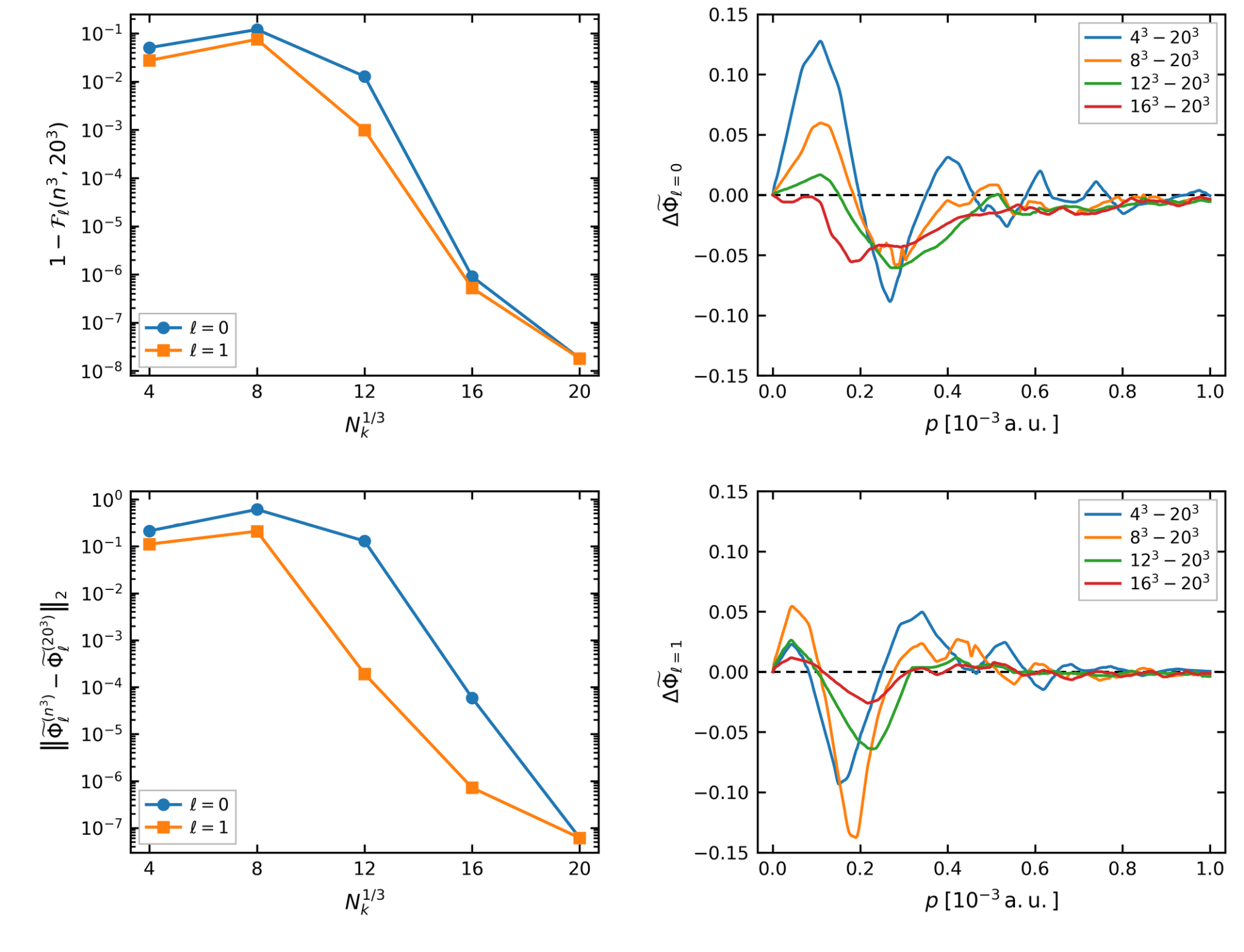}
  \caption{ Stability of the lowest Fredholm eigenfunctions for bulk GaAs relative to   the results of the \(20^3\) mesh. Top left: infidelity , \(1-\mathcal{F}_\ell(n_k^3,20^3)\). Bottom left: phase-aligned Euclidean distance \(\mathcal{D}_\ell(n_k^3,20^3)\). Top right: phase-aligned pointwise differences in the \(\ell=0\) channel for different ... meshes. Bottom right: corresponding differences in the \(\ell=1\) channel. The decreasing global metrics and pointwise residuals demonstrate stabilization of the radial states with increasing mesh density.  }
  \label{fig:gaas_eigenfunction_convergence}
\end{figure*}

\subsection{Radial representation of the screened Coulomb interaction}
\label{sec:gaas_screening}

Figure~\ref{fig:gaas_screened_interaction} shows the mesh dependence of the isotropic real-space screened Coulomb interaction \(W_{\mathrm{cen}}(r)\), reconstructed from the microscopic dielectric response and computed for the Fredholm mesh sequence which is used in Eq.~\eqref{eq:VlBessel}. At short distances, \(W_{\mathrm{cen}}(r)\) exhibits pronounced oscillations. 
Truncating the reciprocal-space representation can introduce oscillations when the interaction is transformed to real space. 
The local-field components of the microscopic dielectric matrix describe variations in screening within the unit cell and modify the short-range interaction. 
The oscillations decrease with distance and approaches a smoother long-range form. The \(4^3\) and \(8^3\) curves remain visibly separated from the denser-mesh results, whereas the \(16^3\) and \(20^3\) curves are substantially closer to each other over the entire radial interval relevant to the transform. This behavior indicates progressive stabilization of both the short-range structure and the long-wavelength screening entering the Fredholm operator.

The long-range part of the screened interaction is particularly important for Wannier-Mott excitons, because their spatial extent is much larger than the lattice spacing. 
This large spatial extent corresponds to a narrow momentum-space envelope, making the binding energy sensitive to screening at small momentum transfers
\(\mathbf{p}-\mathbf{p}'\). 
In bulk GaAs, the relevant band edges lie at \(\mathbf{k}_0=\Gamma\), and the envelope is concentrated near \(\mathbf{p}=0\). 
We therefore examine the numerical stability of \(W_{\mathrm{cen}}(r)\) before assessing the projected kernel, its eigenfunctions, and its lowest eigenvalue.
The numerical assessment follows the sequence
\begin{equation}
  N_k \longrightarrow W_{\mathrm{cen}}(r)
  \longrightarrow V_\ell(p,p')
  \longrightarrow \widetilde{\Phi}_\ell(p), E_B\,,
\end{equation}
so that convergence is tested at the level of the operator, the state, and
the eigenvalue rather than from the final scalar quantity alone.

\subsection{Stability of the radial eigenfunctions}
\label{sec:gaas_eigenfunction_stability}

Eigenvalue stability alone does not establish the stability of the
corresponding state.
We therefore compare the normalized eigenvectors of the symmetric Nystr\"{o}m matrix in Eq.~\eqref{eq:MatrixFredholm} with those obtained on the \(20^3\) reference mesh. For a normalized state \(\widetilde{\Phi}^{(n_k^3)}_\ell\), we define the fidelity and the phase-aligned Euclidean distance as
\begin{align}
  \mathcal{F}_\ell(n_k^3,20^3)
  &=
  \left|
    \left\langle
      \widetilde{\Phi}^{(20^3)}_\ell
      \middle|
      \widetilde{\Phi}^{(n_k^3)}_\ell
    \right\rangle
  \right|^2,
  \\
  \mathcal{D}_\ell(n_k^3,20^3)
  &=
  \min_{\chi}
  \left\|
    \widetilde{\Phi}^{(n_k^3)}_\ell
    -e^{i\chi}\widetilde{\Phi}^{(20^3)}_\ell
  \right\|_2.
  \label{eq:gaas_eigenfunction_metrics}
\end{align}
The minimization over the global phase \(\chi\) removes the arbitrary phase (or sign for a real
eigenvector) and makes the comparison depend only on the eigenfunction shape.

Figure~\ref{fig:gaas_eigenfunction_convergence} shows these two global metrics together with the phase-aligned pointwise differences for the \(\ell=0\) and \(\ell=1\) channels. Both the infidelity and the Euclidean distance decrease as the mesh is refined, and the pointwise residuals become correspondingly smaller. The largest changes occur on the coarse meshes, whereas the \(16^3\) eigenfunctions are already close to the \(20^3\) reference profiles. 
The binding energy and the radial states in both angular-momentum channels show progressively smaller changes as the Brillouin-zone mesh is refined.

\subsection{Convergence of the exciton binding-energy}
\label{sec:gaas_energy_convergence}

Finally, we assess how the binding energy of the lowest Fredholm state obtained from Eq.~\eqref{eq:MatrixFredholm} changes with Brillouin-zone sampling.
For successive meshes \(n_{k,i-1}^3\) and \(n_{k,i}^3\), we evaluate the absolute change
\begin{equation}
  \Delta E_B(n_{k,i}^3,n_{k,i-1}^3)
  =
  \left|E_B(n_{k,i}^3)-E_B(n_{k,i-1}^3)\right|.
  \label{eq:binding_difference_definition}
\end{equation}

\begin{figure}[!t]
  \centering
  \includegraphics[width=\columnwidth]
  {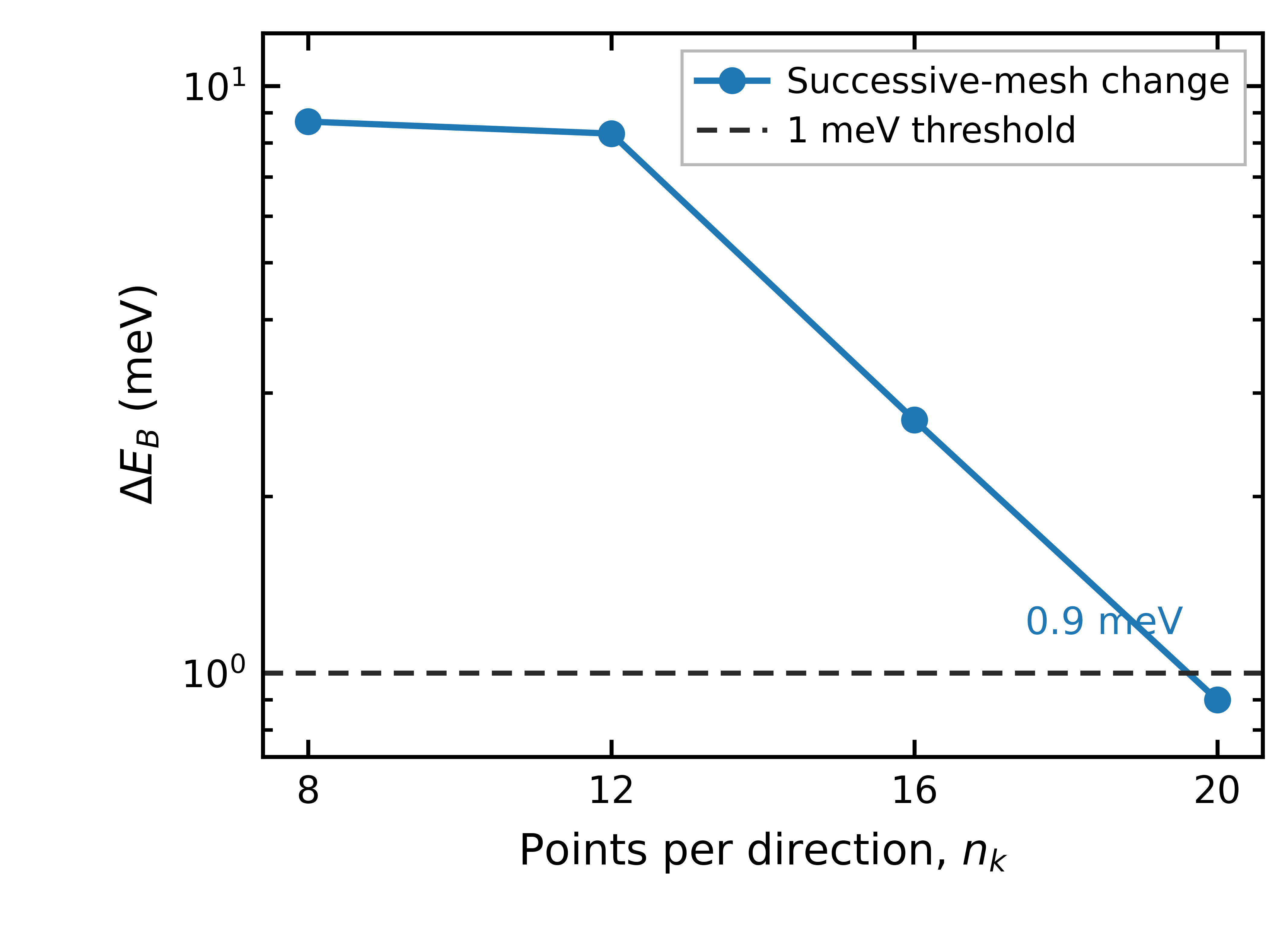}
  \caption{Absolute change in the lowest Fredholm exciton binding energy between successive Brillouin-zone samplings. Each point is plotted at the finer mesh of the corresponding refinement. The horizontal dashed line marks the adopted \(1~\mathrm{meV}\) stability threshold. The \(16^3\!\to20^3\) change is \(0.9~\mathrm{meV}\) and therefore satisfies the criterion in Eq.~\eqref{eq:binding_convergence_criterion}.   }
  \label{fig:gaas_delta_convergence}
\end{figure}

We adopt the explicit finite-mesh stability criterion
\begin{equation}
  \Delta E_B(n_{k,i}^3,n_{k,i-1}^3)<1~\mathrm{meV},
  \label{eq:binding_convergence_criterion}
\end{equation}
which is applied directly to adjacent calculated meshes, \ie without fitting to a
presumed asymptotic law or extrapolation. The successive differences are listed in Table~\ref{tab:gaas_delta_available} and shown in Fig.~\ref{fig:gaas_delta_convergence}. The first two refinements change the binding energy by more than \(8~\mathrm{meV}\), confirming that the coarse meshes do not provide stable eigenvalues. The change then decreases to \(2.7~\mathrm{meV}\) for \(12^3\!\to16^3\) and to \(0.9~\mathrm{meV}\) for \(16^3\!\to20^3\). Thus, the final refinement is the first in the computed sequence to satisfy Eq.~\eqref{eq:binding_convergence_criterion}. The \(20^3\) calculation gives \(E_B\simeq28~\mathrm{meV}\), which we adopt as final result with an operational mesh-stability tolerance of \(1~\mathrm{meV}\).

\begin{table}[!t]
  \caption{
  Absolute change in the lowest Fredholm exciton binding energy between successive Brillouin-zone samplings.   }
  \label{tab:gaas_delta_available}
  \begin{ruledtabular}
    \begin{tabular}{cc}
      Mesh refinement & \(\Delta E_B\) (\(\mathrm{meV}\)) \\
      \hline
      \(4^3\!\to8^3\)   & \(8.7\) \\
      \(8^3\!\to12^3\)  & \(8.3\) \\
      \(12^3\!\to16^3\) & \(2.7\) \\
      \(16^3\!\to20^3\) & \(0.9\) \\
    \end{tabular}
  \end{ruledtabular}
\end{table}

The screened Coulomb interaction and radial eigenfunctions change only slightly between the \(16\times16\times16\) and \(20\times20\times20\) Brillouin-zone meshes. 
Over the same refinement, the binding energy changes by \(0.9~\mathrm{meV}\), as shown in Fig.~\ref{fig:gaas_delta_convergence}.
This change is below the adopted \(1~\mathrm{meV}\) stability threshold and quantifies the remaining sampling dependence between the two finest meshes considered.

\section{Conclusions}
\label{sec:conc}

We have developed a radial Fredholm formulation of the solid-state
BSE for Wannier-Mott excitons in direct-gap bulk materials.
Within the static-kernel and Tamm-Dancoff approximations, it
describes relative electron-hole motion using a single
valence-conduction channel, material-dependent effective masses,
and a central projection of the microscopic screened Coulomb
interaction. This reduction is particularly well suited to
spatially extended excitons when transitions near the band extrema
dominate and coupling to other band pairs is weak. The model
retains the screened direct attraction and omits bare electron-hole
exchange, leaving exchange-induced singlet-triplet splitting
outside its scope. The conventional triplet matrix BSE therefore
provides the closest like-for-like reference. Indirect excitons
and noncentral interactions would require extensions to finite
center-of-mass momentum and coupled angular-momentum channels,
respectively.

The computational advantage is that the exciton matrix dimension
is controlled by the radial grid independently of the
Brillouin-zone sampling used for screening. Denser screening
meshes therefore do not require larger radial matrices.
The eigenvalue problems are solved deterministically, giving
direct access to binding energies and radial states without
stochastic sampling uncertainty. Once the electronic-structure
and screening inputs are available, this compact formulation
offers a route to efficient preliminary screening of candidate
materials through their Wannier-Mott exciton properties.
Its predictive accuracy depends on the validity of the stated
physical approximations.

For bulk GaAs, the lowest Fredholm eigenvalue agrees with both
the singlet and triplet matrix-BSE values within \(1~\mathrm{meV}\)
on the finest shared mesh. Further refinement stabilizes the
screened Coulomb interaction and radial states, with the finest
calculation satisfying the adopted finite-mesh stability
criterion for the binding energy.

\begin{acknowledgments}
This work was carried out at the Department of Physics and the Center
for the Science of Materials Berlin, Humboldt-Universit\"at zu Berlin,
12489 Berlin, Germany.
Ch.T. acknowledges support from the Alexander von Humboldt Foundation
through a Humboldt Research Fellowship for Experienced Researchers
(Ref. No.~3.3-RUS-1214771-HFST-E).
Ch.T. acknowledges ISCRA for awarding this project access to the LEONARDO
supercomputer, owned by the EuroHPC Joint Undertaking and hosted by
CINECA (Italy), under the Class B allocation
(Project No.~HP10BHHWLK).
\end{acknowledgments}


\appendix

\section{Construction of the central screened interaction}
\label{sec:SM_central}

The Fredholm reduction uses the translationally invariant central component
of the static microscopic screened interaction.  In a plane-wave
representation, the full two-point interaction is
\begin{align}
W(\mathbf r,\mathbf r')
&=
\frac{1}{N_q\Omega}
\sum_{\mathbf q}
\sum_{\mathbf G,\mathbf G'}
e^{i(\mathbf q+\mathbf G)\cdot\mathbf r}
W_{\mathbf q}(\mathbf G,\mathbf G')
e^{-i(\mathbf q+\mathbf G')\cdot\mathbf r'},
\label{eq:screening_RS_FT}
\end{align}
where \(N_q\) is the number of sampling points, \(\Omega\) is the
primitive-cell volume, and \(W_{\mathbf q}(\mathbf G,\mathbf G')\) is formed
from the full microscopic dielectric matrix in \exciting.

We introduce the common coordinate
\(\mathbf X=(\mathbf r+\mathbf r')/2\) and the relative coordinate
\(\mathbf s=\mathbf r-\mathbf r'\).  Averaging over \(\mathbf X\) within one
primitive cell gives
\begin{align}
\overline W(\mathbf s)
&=
\frac{1}{\Omega}
\int_{\Omega}d^3X\,
W\!\left(\mathbf X+\frac{\mathbf s}{2},
          \mathbf X-\frac{\mathbf s}{2}\right)
\nonumber\\
&=
\frac{1}{N_q\Omega^2}
\sum_{\mathbf q}
\sum_{\mathbf G,\mathbf G'}
W_{\mathbf q}(\mathbf G,\mathbf G')
e^{i[\mathbf q+(\mathbf G+\mathbf G')/2]\cdot\mathbf s} \nonumber \\
&\int_{\Omega}d^3X\,
e^{i(\mathbf G-\mathbf G')\cdot\mathbf X}
\nonumber\\
&=
\frac{1}{N_q\Omega}
\sum_{\mathbf q}\sum_{\mathbf G}
e^{i(\mathbf q+\mathbf G)\cdot\mathbf s}
W_{\mathbf q}(\mathbf G,\mathbf G).
\label{eq:SM_center_average}
\end{align}
The final spherical average defines the interaction used in the radial
equations,
\begin{align}
W_{\mathrm{cen}}(r)
&=
\frac{1}{4\pi}
\int d\Omega_{\hat{\mathbf s}}\,
\overline W(r\hat{\mathbf s})
\nonumber\\
&=
\frac{1}{N_q\Omega}
\sum_{\mathbf q}\sum_{\mathbf G}
W_{\mathbf q}(\mathbf G,\mathbf G)
j_0\!\left(|\mathbf q+\mathbf G|r\right)
\nonumber\\
&=
\frac{1}{N_q\Omega}
\sum_{\mathbf q}\sum_{\mathbf G}
W_{\mathbf q}(\mathbf G,\mathbf G)
\frac{\sin(|\mathbf q+\mathbf G|r)}{|\mathbf q+\mathbf G|r}.
\label{eq:SM_Wcen}
\end{align}
The full dielectric matrix is inverted before this projection.  Consequently,
off-diagonal local-field couplings affect the retained diagonal elements
\(W_{\mathbf q}(\mathbf G,\mathbf G)\), even though the explicit
\(\mathbf G\ne\mathbf G'\) components, including the wings, vanish under the
common-coordinate average.
The head is retained. The screening is computed on the symmetrized
momentum-transfer grid \(\mathbf q=\mathbf k-\mathbf k'\).
The common offset of the \(\mathbf k\)-mesh cancels in this
difference, so the \(\mathbf q\)-grid includes \(\mathbf q=0\).
The singular Coulomb contribution is estimated by spherical
averaging in reciprocal space around \(\mathbf q=0\).
The full screening meshes contain \(N_q=n_k^3\) points, with
\(n_k\in\{4,8,12,16,20\}\).

In this notation, \(W_{\mathrm{cen}}\) is the screened Coulomb interaction
without an attractive minus sign.  The electron--hole attraction is introduced
explicitly in the direct BSE kernel and in the radial Hamiltonian, as in the
main text.

\section{Radial cutoff and finite-section control}
\label{sec:SM_Lambda}

This section gives the rationale for the radial momentum cutoff \(\Lambda\),
the finite-section and tail diagnostics, and the numerical parameters used in
the GaAs calculations.  The cutoff must be large enough to contain the
momentum-space exciton envelope, but it must also remain within the range over
which the effective-mass approximation is valid.  Pointwise decay of
\(V_\ell(p,p')\) is not an appropriate criterion because the spherical-Bessel
kernel is oscillatory.  Unless stated otherwise, atomic units are used below.

\subsection{Radial Fredholm equation and finite-section truncation}
\label{sec:SM_finite_section}

For each angular momentum \(\ell\), the main text yields
\begin{align}
\left(\frac{p^2}{2\mu}+E_B\right)\Phi_\ell(p)
&=
\frac{4\pi}{(2\pi)^3}
\int_{0}^{\infty}p'^2\,dp'\,
V_\ell(p,p')\,\Phi_\ell(p') ,
\label{eq:SM_radial_infinite}
\end{align}
with
\begin{align}
V_\ell(p,p')
&=
4\pi\int_{0}^{\infty}dr\,r^2\,
j_\ell(pr)\,j_\ell(p'r)\,W_{\mathrm{cen}}(r).
\label{eq:SM_Vl_bessel}
\end{align}
Thus, \(V_\ell\) is defined without an attractive minus sign.  The associated
Hamiltonian is the kinetic operator minus the integral operator below.

Numerically we solve a \emph{finite-section} problem: truncate momentum to $p\in[0,\Lambda]$ and define the truncated operator
\begin{align}
\left(\mathcal{V}_\ell^{(\Lambda)}\Phi\right)(p)
:=
\frac{4\pi}{(2\pi)^3}
\int_{0}^{\Lambda}p'^2\,dp'\,
V_\ell(p,p')\,\Phi(p'),
\label{eq:SM_V_Lambda_def}
\end{align}
with $0\le p\le \Lambda$,
so that
\begin{align}
\left(\frac{p^2}{2\mu}+E_B^{(\Lambda)}\right)\Phi_\ell^{(\Lambda)}(p)
&=
\left(\mathcal{V}_\ell^{(\Lambda)}\Phi_\ell^{(\Lambda)}\right)(p).
\label{eq:SM_radial_truncated}
\end{align}

The truncation removes the tail $p'>\Lambda$ in the integral operator.
Acting on the \emph{exact} solution $\Phi_\ell$, the omitted tail contribution reads
\begin{align}
\left(\mathcal{V}_\ell-\mathcal{V}_\ell^{(\Lambda)}\right)\Phi_\ell
&=
\frac{4\pi}{(2\pi)^3}
\int_{\Lambda}^{\infty}p'^2\,dp'\,
V_\ell(p,p')\,\Phi_\ell(p') .
\label{eq:SM_tail_operator_error}
\end{align}
Therefore, the truncation is controlled not by pointwise smallness of $V_\ell(p,p')$, but by the \emph{weighted tail} $p'^2\Phi_\ell(p')$ beyond $\Lambda$ being negligible in
norm.
This motivates the tail diagnostics introduced in Sec.~\ref{sec:SM_protocol}.

\subsection{Nonmonotonic structure of the radial kernel}
\label{sec:SM_oscillations}

The integrand in Eq.~\eqref{eq:SM_Vl_bessel} contains spherical Bessel functions.
For large argument $z$ one has the oscillatory limiting form
\begin{align}
j_\ell(z)
&=
\frac{1}{z}\sin\!\left(z-\frac{\ell\pi}{2}\right)
+\mathcal{O}\!\left(z^{-2}\right),
\qquad z\to\infty,
\label{eq:SM_jl_asymptotic}
\end{align}
so $j_\ell(pr)j_\ell(p'r)$ oscillates rapidly in $r$ with a slowly decaying envelope.
Consequently, $V_\ell(p,p')$ can show oscillatory structure as a function of $p$ and
$p'$ and is not expected to decay monotonically.
Equation~\eqref{eq:SM_jl_asymptotic} explains why pointwise ``kernel decay''
is an unreliable cutoff criterion; the finite section must instead be
assessed through its spectrum and the boundary weight of its
eigenvectors.\cite{DLMF10}

A useful heuristic for the momentum-space tail of the \emph{envelope} follows from balancing the kinetic term with a screened kernel whose large-momentum transfer behaves as $\sim 1/q^2$ (explicitly verified for Yukawa below).
For \(\ell=0\), the magnitude behaves schematically as
\begin{align}
\frac{p^2}{2\mu}\,\Phi_0(p)
&\approx
\frac{\mathcal C_0}{p^2},
\label{eq:SM_tail_balance}
\\
\Rightarrow\qquad
|\Phi_0(p)|
&\sim
\frac{2\mu|\mathcal C_0|}{p^4},
\qquad p\to\infty,
\label{eq:SM_tail_phi_p4}
\end{align}
so the norm tail $\int_\Lambda^\infty dp\,p^2|\Phi_0(p)|^2$ decays rapidly once $\Lambda$ exceeds the dominant physical momentum scale.
For fixed $p'$, $V_\ell(p,p')$ exhibits a pronounced ridge/maximum around $p\simeq p'$, reflecting the dominance of small momentum transfer $q=|\mathbf{p}-\mathbf{p}'|$ rather than a monotone decay in either argument.
In a two-dimensional map over \((p,p')\), the same structure appears as a ridge along $p\simeq p'$, confirming that it is generally not meaningful to choose the momentum cutoff $\Lambda$ by demanding pointwise smallness of $V_\ell(p,p')$ at $p=\Lambda$, but instead by convergence of spectral observables (binding energies) and
wave-function tail diagnostics, as discussed in Sec.~\ref{sec:SM_protocol}.

\subsection{Analytic Yukawa test}
\label{sec:SM_yukawa}

For validation we use the positive screened-interaction magnitude
\begin{align}
W_{\mathrm{cen}}(r)
&=
A\,\frac{e^{-\kappa r}}{r},
\qquad A>0,\;\kappa>0 .
\label{eq:SM_yukawa_realspace}
\end{align}
Its (3D) Fourier transform reads
\begin{align}
\widetilde{W}(q)
&=
\frac{4\pi A}{q^2+\kappa^2},
\qquad q=|\mathbf{q}|,
\label{eq:SM_yukawa_momentum}
\end{align}
which is used as a numerical implementation check.

In the isotropic Wannier limit the direct kernel depends on the momentum transfer
$\mathbf{q}=\mathbf{p}-\mathbf{p}'$, so
\begin{align}
\widetilde W(\mathbf{p}-\mathbf{p}')
&=
\widetilde{W}(|\mathbf{p}-\mathbf{p}'|)
=
\frac{4\pi A}{p^2+p'^2-2pp'\cos\theta+\kappa^2},
\label{eq:SM_yukawa_kernel_angle}
\end{align}
with $\cos\theta=\hat{\mathbf{p}}\cdot\hat{\mathbf{p}}'.$
The partial-wave component is the Legendre projection
\begin{align}
V_\ell(p,p')
&=
\frac{1}{2}\int_{-1}^{1}dx\,P_\ell(x)\,
\widetilde W(\mathbf{p}-\mathbf{p}')
\Big|_{x=\cos\theta}
\notag\\
&=
2\pi A\int_{-1}^{1}dx\,
\frac{P_\ell(x)}{p^2+p'^2+\kappa^2-2pp'x}.
\label{eq:SM_yukawa_projection_start}
\end{align}
Introduce
\begin{align}
z
&:=
\frac{p^2+p'^2+\kappa^2}{2pp'}
\,\Longrightarrow\,
p^2+p'^2+\kappa^2-2pp'x
=
2pp'(z-x),
\label{eq:SM_z_def}
\end{align}
so
\begin{align}
V_\ell(p,p')
&=
\frac{\pi A}{pp'}\int_{-1}^{1}dx\,\frac{P_\ell(x)}{z-x}.
\label{eq:SM_yukawa_projection_mid}
\end{align}
Using Neumann's integral representation of the Legendre function of the second kind,
\begin{align}
Q_\ell(z)
&=
\frac{1}{2}\int_{-1}^{1}dx\,\frac{P_\ell(x)}{z-x},
\qquad z>1,
\label{eq:SM_neumann_Q}
\end{align}
we obtain the closed form
\begin{align}
V_\ell(p,p')
&=
\frac{2\pi A}{pp'}\,Q_\ell\!\left(\frac{p^2+p'^2+\kappa^2}{2pp'}\right).
\label{eq:SM_yukawa_Vl_closed}
\end{align}
Equations~\eqref{eq:SM_neumann_Q}--\eqref{eq:SM_yukawa_Vl_closed}
provide a reference for validating the numerical evaluation of
Eq.~\eqref{eq:SM_Vl_bessel}.\cite{DLMF10}

\subsection{Finite-section parameters used for GaAs}
\label{sec:SM_abinitio}

For GaAs, \(W_{\mathrm{cen}}(r)\) is constructed directly from the microscopic
screened interaction using Eq.~\eqref{eq:SM_Wcen}; no Yukawa fit is used in
the production calculation.  The radial discretization parameters are listed
in Table~\ref{tab:SM_radial_parameters}.  They are held fixed throughout the
\(N_k=N_q=n_k^3\) sequence, so that the changes reported in the main text probe
Brillouin-zone sampling rather than a simultaneous change of the radial
finite section.

\begin{table}[!t]
  \caption{Fixed parameters of the GaAs radial Fredholm calculation.}
  \label{tab:SM_radial_parameters}
  \begin{ruledtabular}
    \begin{tabular}{lc}
      Quantity & Value \\
      \hline
      Maximum angular momentum, \(\ell_{\max}\) & \(1\) \\
      Momentum cutoff, \(\Lambda\) & \(0.020~\mathrm{bohr}^{-1}\) \\
      Number of radial momentum points, \(N_p\) & \(400\) \\
      Auxiliary momentum-cutoff factor, \(p_{\max}^{\mathrm{factor}}\) & \(1.05\) \\
      Real-space cutoff, \(r_{\max}\) & \(4000~\mathrm{bohr}\) \\
      Number of real-space radial points, \(N_r\) & \(3000\) \\
      Real-space tail-truncation parameter & \(0.7\) \\
    \end{tabular}
  \end{ruledtabular}
\end{table}

The reduced mass is obtained from the quasiparticle dispersion around
\(\Gamma\), using three points along each reciprocal-space direction with
spacing \(\Delta k=0.055\) in the reciprocal-coordinate convention of the
\exciting\ calculation.  The momentum interval is restricted to the
near-edge region described by this effective-mass model; extending the
finite section beyond that validity window would not improve the underlying
one-valence-band/one-conduction-band approximation.

\subsection{Nystr\"om discretization and the quantities that must converge}
\label{sec:SM_nystrom}

On $[0,\Lambda]$ we discretize by a quadrature rule
\begin{align}
\int_{0}^{\Lambda} f(p')\,dp'
&\approx
\sum_{j=1}^{N_p} w_j\,f(p_j),
\label{eq:SM_quadrature}
\end{align}
leading to the symmetrized Nystr\"om eigenproblem for a state
\(\lambda\) in channel \(\ell\)
\begin{align}
\sum_{j=1}^{N_p}
\tilde{H}^{(\ell)}_{ij}\,\tilde{\Phi}_{\ell,j}
&=
\mathcal{E}_{\lambda}\,\tilde{\Phi}_{\ell,i},
\label{eq:SM_discrete_eig}
\\
\tilde{H}^{(\ell)}_{ij}
&=
\delta_{ij}\,\frac{p_i^2}{2\mu}
-
\frac{4\pi}{(2\pi)^3}
\sqrt{w_i w_j}\,p_i p_j\,
V_\ell(p_i,p_j),
\label{eq:SM_H_sym}
\\
E_B^{(\Lambda)}
&=
-\mathcal{E}_{\lambda} \quad (\mathcal{E}_{\lambda}<0).
\label{eq:SM_EB_from_lambda}
\end{align}
The matrix is real symmetric when
\(V_\ell(p_i,p_j)=V_\ell(p_j,p_i)\).  This is the same sign convention and
normalization used in the main text.  Nystr\"om approximation of compact
integral operators is standard.\cite{KressLIE,AtkinsonSIAM1967}

\subsection{Finite-section and boundary-tail diagnostics}
\label{sec:SM_protocol}

For a trial sequence of cutoffs, the finite-section change of a targeted
binding energy is
\begin{align}
\Delta_\Lambda E_B^{\lambda}
&=
\left|E_B^{\lambda}(\Lambda_{n+1})-E_B^{\lambda}(\Lambda_n)\right|.
\label{eq:SM_deltaE}
\end{align}
For a chosen absolute working tolerance \(\tau_E\), finite-section stability
requires \(\Delta_\Lambda E_B^{\lambda}<\tau_E\).  In addition, because the omitted
contribution in Eq.~\eqref{eq:SM_tail_operator_error} is weighted by the
eigenfunction, the fraction of the discrete norm in the outer part of the
momentum interval provides a boundary diagnostic.  With
\(\widetilde{\Phi}_{\ell,i}=\sqrt{w_i}\,p_i\Phi_\ell(p_i)\), this fraction is
\begin{align}
\eta_\beta^{\lambda}(\Lambda)
&:=
\frac{\displaystyle
      \sum_{p_i\ge\beta\Lambda}
      \left|\widetilde{\Phi}^{\lambda}_{\ell,i}\right|^2}
     {\displaystyle
      \sum_{i=1}^{N_p}
      \left|\widetilde{\Phi}^{\lambda}_{\ell,i}\right|^2},
\qquad 0<\beta<1.
\label{eq:SM_eta_tail}
\end{align}
The cutoff should be increased if either the eigenvalue remains sensitive to
the finite section or \(\eta_\beta^{\lambda}\) shows that the state is concentrated
near the numerical boundary.  This is a boundary diagnostic, not an
extrapolation to \(\Lambda\to\infty\).  The operational
\(1~\mathrm{meV}\) criterion reported in the main text concerns consecutive
Brillouin-zone meshes at the fixed radial parameters of
Table~\ref{tab:SM_radial_parameters}; it is not an infinite-mesh
extrapolation.

\subsection{Control of the real-space quadrature}
\label{sec:SM_r_quadrature}

When $V_\ell(p,p')$ is evaluated via Eq.~\eqref{eq:SM_Vl_bessel}, the $r$-quadrature must
resolve the fastest oscillations at $p,p'\lesssim\Lambda$. From Eq.~\eqref{eq:SM_jl_asymptotic}
the oscillation period is $\Delta r \sim \pi/p$. A conservative resolution criterion is
\begin{align}
\Delta r
&\lesssim
\frac{\pi}{M_{\mathrm{osc}}\,\Lambda},
\qquad M_{\mathrm{osc}}\simeq 10.
\label{eq:SM_dr_rule}
\end{align}
Here \(M_{\mathrm{osc}}\) is a dimensionless resolution factor.
The production values \(r_{\max}=4000~\mathrm{bohr}\), \(N_r=3000\), and
\(\Lambda=0.020~\mathrm{bohr}^{-1}\) are reported in
Table~\ref{tab:SM_radial_parameters}.  The analytic Yukawa result in
Eq.~\eqref{eq:SM_yukawa_Vl_closed} provides an independent check of the
finite-\(r\) quadrature.

\begin{figure}[!t]
  \centering
  \includegraphics[width=1.0\linewidth]{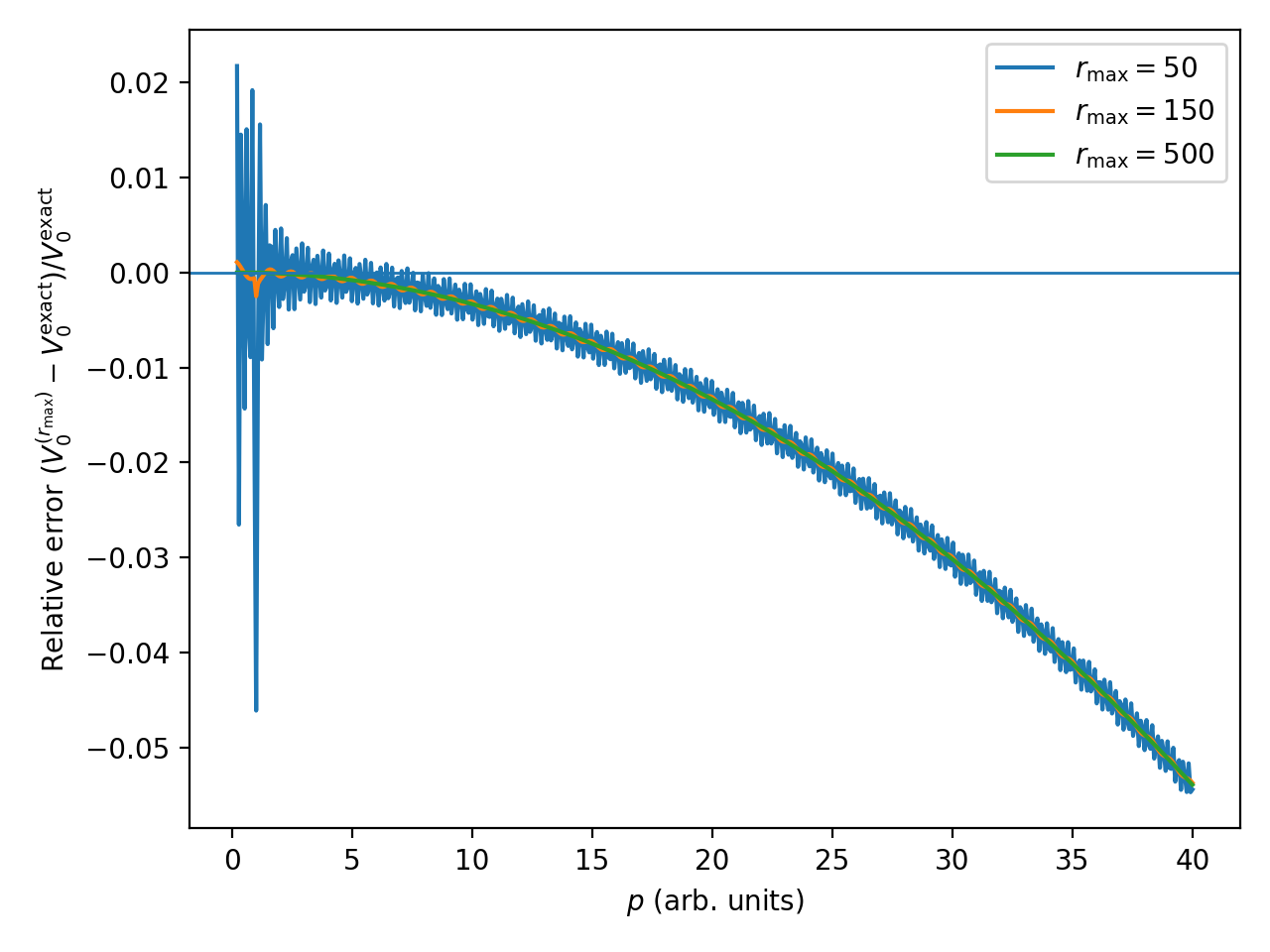}
  \caption{Relative finite-\(r_{\max}\) error in the positive Yukawa kernel,
  \((V^{(r_{\max})}_0-V^{\mathrm{exact}}_0)/V^{\mathrm{exact}}_0\), as a
  function of \(p\) at fixed \(p'=1\).  The truncated real-space integrals use
  \(r_{\max}\in\{50,150,500\}\), a uniform radial spacing, \(A=1\), and
  \(\kappa=0.02\) (arbitrary units).  Increasing \(r_{\max}\) suppresses the
  spurious oscillations caused by truncating the Bessel integral.}
  \label{fig:SM_Vl_trunc}
\end{figure}

Figure~\ref{fig:SM_Vl_trunc} demonstrates a related numerical point: when $V_\ell(p,p')$
is evaluated via the oscillatory Bessel integral \eqref{eq:SM_Vl_bessel} with a finite
upper limit $r_{\max}$, truncation induces nonphysical oscillations as a function of $p$.
These artifacts decrease systematically as $r_{\max}$ is increased, motivating an
independent convergence check of the $r$-quadrature (and $r_{\max}$) at the largest
momenta $p,p'\lesssim \Lambda$ so that apparent $\Lambda$-dependence is not contaminated
by kernel-evaluation error.

\section{Numerical construction of the radial kernels}
\label{sec:FastKernels}

The expensive step in the reduced calculation is the construction of
\(V_\ell(p_i,p_j)\), rather than the diagonalization of the final
\(N_p\times N_p\) matrix.  Once \(W_{\mathrm{cen}}(r)\) has been evaluated
from Eq.~\eqref{eq:SM_Wcen}, the calculation uses the same radial definition
as the main text,
\begin{align}
V_\ell(p_i,p_j)
&=
4\pi\int_0^{r_{\max}}dr\,r^2\,
j_\ell(p_i r)j_\ell(p_j r)W_{\mathrm{cen}}(r).
\label{eq:FastKernel_Vl_def}
\end{align}
Let \(r_\alpha\) and \(w_\alpha^{(r)}\) denote the real-space nodes and quadrature
weights, respectively.  For each retained angular momentum we tabulate
\begin{align}
B^{(\ell)}_{\alpha i}
&=
j_\ell(p_i r_\alpha),
\label{eq:FastKernel_B_matrix}
\\
d_\alpha
&=
4\pi w_\alpha^{(r)} r_\alpha^2W_{\mathrm{cen}}(r_\alpha).
\label{eq:FastKernel_D_weights}
\end{align}
The kernel is then assembled as
\begin{align}
V_\ell(p_i,p_j)
&\simeq
\sum_{\alpha=1}^{N_r}
B^{(\ell)}_{\alpha i}\,
d_\alpha\,
B^{(\ell)}_{\alpha j},
\label{eq:FastKernel_discrete_realspace}
\\
\mathbf V_\ell
&\simeq
\bigl(\mathbf B^{(\ell)}\bigr)^{\mathsf T}
\mathbf C\,
\mathbf B^{(\ell)},
\qquad
\mathbf C=\operatorname{diag}(d_1,\ldots,d_{N_r}).
\label{eq:FastKernel_gram}
\end{align}
This weighted Gram construction evaluates every spherical Bessel factor once
per \((r_\alpha,p_i,\ell)\) and reuses it for all matrix elements.  Because
the radial grids and the retained values \(\ell=0,1\) are common to the full
Brillouin-zone mesh sequence, the Bessel tables are also reused when only
\(W_{\mathrm{cen}}(r_\alpha)\) changes.

Equation~\eqref{eq:FastKernel_gram} is symmetric by construction, and that
symmetry is retained by the Nystr\"om rescaling.  The momentum nodes are uniformly spaced on
\(0\le p_i\le\Lambda\), with corresponding quadrature weights \(w_i\).  The numbers of nodes and the interval
boundaries are those in Table~\ref{tab:SM_radial_parameters}.

For reference, the relation
\begin{align}
j_\ell(x)
&=
\sqrt{\frac{\pi}{2x}}J_{\ell+1/2}(x)
\label{eq:FastKernel_jl_to_Jnu}
\end{align}
shows that Eq.~\eqref{eq:FastKernel_Vl_def} contains two Bessel functions of
half-integer order.\cite{DLMF10}  Its oscillatory
finite-domain error is controlled through the \(r_{\max}\) test illustrated
in Fig.~\ref{fig:SM_Vl_trunc}, while the analytic Yukawa expression
\eqref{eq:SM_yukawa_Vl_closed} checks the normalization and sign.

\section{Implementation in \textsc{exciting} and relation to the matrix BSE}
\label{sec:SM_implementation}

The conventional calculation in \exciting\ constructs the BSE Hamiltonian in
the Bloch-transition basis \(\ket{vc\mathbf k}\).  The Fredholm implementation
uses the same underlying quasiparticle and dielectric data but retains one
valence and one conduction band in the effective-mass reduction and applies
the central projection in Appendix~\ref{sec:SM_central}.  The implementation used for the
calculations reported in the main text follows these steps:

\begin{enumerate}
  \item read the quasiparticle band edges and dispersions around the direct
        gap and determine the reduced mass \(\mu\);
  \item construct the microscopic static screened interaction from the full
        dielectric matrix and form \(W_{\mathrm{cen}}(r)\) using
        Eq.~\eqref{eq:SM_Wcen};
  \item generate the radial \(p\)- and \(r\)-space grids, tabulate the
        spherical Bessel functions, and assemble
        \(V_\ell(p_i,p_j)\) with Eq.~\eqref{eq:FastKernel_gram};
  \item form the real-symmetric Nystr\"om Hamiltonian
        \(\widetilde{\mathbf H}^{(\ell)}\) and compute its low-lying
        eigenpairs;
  \item obtain \(E_B^{\lambda}=-\mathcal{E}_{\lambda}\) for every negative eigenvalue and retain
        the normalized radial envelopes for the state-convergence tests.
\end{enumerate}

For the GaAs validation, the conventional singlet and triplet matrix-BSE
calculations and the Fredholm calculations use common \(4^3\), \(8^3\), and
\(12^3\) Brillouin-zone meshes.  These shared meshes test fixed-sampling
consistency.  The Fredholm sequence is then continued to \(16^3\) and
\(20^3\) with the radial parameters in
Table~\ref{tab:SM_radial_parameters} held fixed.  Because the reduced
Hamiltonian contains the screened direct attraction but omits bare
electron--hole exchange, the triplet matrix BSE is the closest like-for-like
reference.  Agreement with the singlet and triplet values at \(12^3\) is
therefore a validation of the reduced implementation at a common finite
sampling, not a claim that the two Hamiltonians are formally identical.

The radial formulation replaces the Bloch-transition Hamiltonian by one
\(N_p\times N_p\) real-symmetric matrix for each retained \(\ell\).  This
dimensional reduction is applicable only under the effective-mass and central
interaction approximations stated in the main text; no corresponding claim is
made for indirect-gap systems or for general noncentral kernels.

\bibliography{main}

\end{document}